\documentclass[sigconf]{acmart}
\usepackage[noend]{algpseudocode}
\usepackage{algorithmicx,algorithm}
\usepackage{url}
\usepackage{multirow}
\usepackage{todonotes}
\usepackage{xcolor}

\usepackage{xspace}

\def\ie{\textit{i.e.}\xspace}
\def\etal{\textit{et al.}\xspace}
\def\etc{\textit{etc.}\xspace}
\def\eg{\textit{e.g.}\xspace}

\AtBeginDocument{%
  \providecommand\BibTeX{{%
    \normalfont B\kern-0.5em{\scshape i\kern-0.25em b}\kern-0.8em\TeX}}}

\setcopyright{acmcopyright}

\copyrightyear{2023}
\acmYear{2023}
\acmConference[Conference '23]{Proceedings}{2023}{Location}

\begin{document}

\title{Cognitive Exoskeleton: Augmenting Human Cognition with an AI-Mediated Intelligent Visual Feedback}

\author{Songlin Xu}
\email{soxu@ucsd.edu}
\orcid{0000-0002-3674-922X}
\affiliation{%
  \institution{University of California San Diego}
  \streetaddress{9500 Gilman Dr, La Jolla}
  \city{San Diego}
  \state{CA}
  \country{USA}
  \postcode{92093}
}

\author{Xinyu Zhang}
\email{xyzhang@ucsd.edu}
\orcid{0000-0001-9688-8056}
\affiliation{%
  \institution{University of California San Diego}
  \streetaddress{9500 Gilman Dr, La Jolla}
  \city{San Diego}
  \state{CA}
  \country{USA}
  \postcode{92093}
}

\renewcommand{\shortauthors}{Xu, et al.}

\begin{abstract}
In this paper, we introduce an AI-mediated framework that can provide intelligent feedback to augment human cognition. Specifically, we leverage deep reinforcement learning (DRL) to provide adaptive time pressure feedback to improve user performance in a math arithmetic task. Time pressure feedback could either improve or deteriorate user performance by regulating user attention and anxiety. Adaptive time pressure feedback controlled by a DRL policy according to users' real-time performance could potentially solve this trade-off problem. However, the DRL training and hyperparameter tuning may require large amounts of data and iterative user studies. Therefore, we propose a dual-DRL framework that trains a regulation DRL agent to regulate user performance by interacting with another simulation DRL agent that mimics user cognition behaviors from an existing dataset. Our user study demonstrates the feasibility and effectiveness of the dual-DRL framework in augmenting user performance, in comparison to the baseline group.

\end{abstract}



\begin{teaserfigure}
  \includegraphics[width=\textwidth]{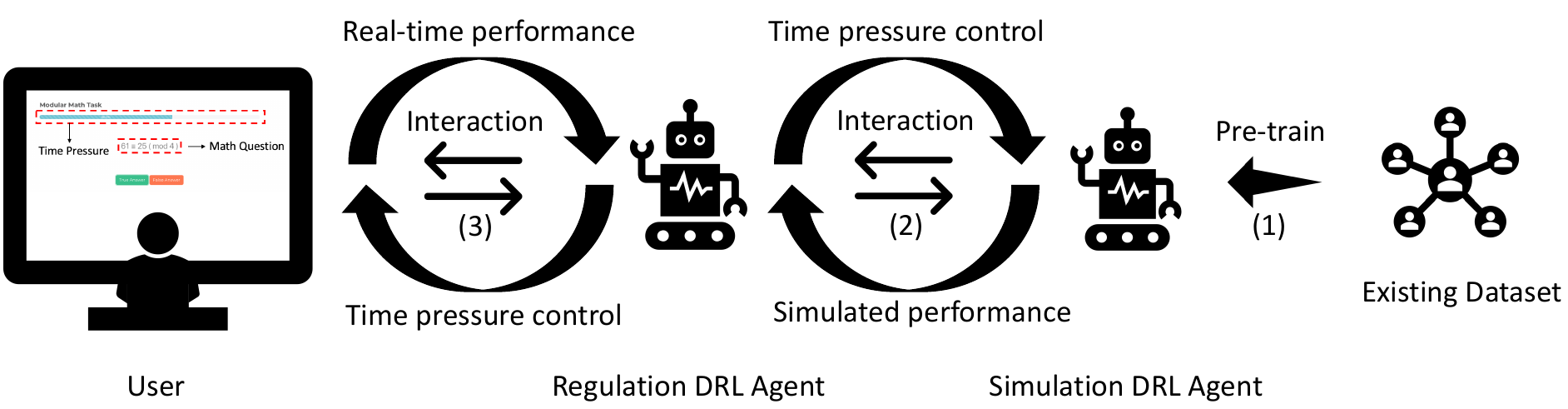}
  \caption{Scenario demonstration of our dual DRL agents framework. The simulation DRL agent is first pre-trained from an existing dataset (1) to mimic human cognition behaviors. The regulation DRL agent could then be trained in the virtual environment by interacting with the simulation DRL agent (2) to learn potential time pressure control strategies for human cognition performance augmentation. The trained regulation DRL agent is then applied on real users (3) to improve their cognition performance using adaptive time pressure control strategies according to users' real-time performance.}
  \Description{Scenario demonstration of our dual DRL agents framework. The simulation DRL agent is first pre-trained from existing dataset (1) to mimic human cognition behaviors. The regulation DRL agent could then be trained in the virtual environment by interacting with the simulation DRL agent (2) to learn potential time pressure control strategies for human cognition performance augmentation. The trained regulation DRL agent is then applied on real users (3) to improve their cognition performance using adaptive time pressure control strategies according to users' real-time performance.}
  \label{teaser}
\end{teaserfigure}

\maketitle

\section{Introduction}
\label{sec:intro}

In spite of the giant leap-forward of AI technologies in the past decade, the human cognitive ability still far eludes machine systems. The situation will likely remain for long, since the current wave of AI resolution heavily relies on massive data, which differs from human cognition in principle, and is unlikely to surpass human's ability of generalization and logical reasoning.  
To fundamentally overcome this limitation, AI can augment or even integrate with humans, instead of a replacement \cite{hbr}. 
One way to attain this vision is to develop intelligent feedback to improve human cognition performance. For example, by properly visualizing brain wave patterns to human users as a neurofeedback, the users could self-regulate their emotion status \cite{faller2019regulation}. 
Such neurofeedback needs to be judiciously presented, as a constant feedback may increase anxiety and reduce cognitive performance. 
Recent research (\eg, AttentivU\cite{kosmyna2019attentivu} and Dan \etal \cite{10.1145/2207676.2207679}) proposed to adapt the feedback over time to improve user engagement. When the system detects a drop of user engagement, it triggers corresponding feedback to regulate users immediately. However, this is a myopic feedback control strategy, and may not improve the overall performance across a cognitive process. Therefore, intelligent feedback to regulate user performance is a general problem that still needs further research.

In this paper, we seek to answer a general question in realizing human-AI integrated systems, \ie 
\textit{can we leverage AI models to augment and improve human cognition performance, through intelligent and continuous feedback?}
We explore the answers through a representative cognitive task, \ie, math
arithmetic problem solving (Sec.~\ref{sec:task}), which is widely used in
evaluation of human cognition
behaviors\cite{lin2011spatial,judd2021training,daitch2016mapping,costa2019boostmeup}.
We use \textit{time pressure} as the feedback modality. 
Time pressure, presented through a visualization of time passing, has been proven to be effective in regulating cognition performance\cite{cheng2017evaluation,slobounov2000neurophysiological,moore2012time}. It 
can increase the arousal of users\cite{edland1993judgment} by delivering a
sense of urgency. However, a too strong time pressure may distract and stress users, thus degrading their performance. Existing
research in cognitive science has demonstrated this trade-off \cite{zur1981effect,slobounov2000neurophysiological}, 
but to our knowledge, no systematic solutions have been proposed and verified to adapt time pressure in real-time to continuously regulate human cognition. 
Ideally, an adaptive algorithm to
control the existence of time pressure feedback according to users' real-time
performance can attain a sweet-spot. 
However, it is challenging to design such adaptation, as human 
cognition behaviors are highly dynamic and, unlike machine systems, do not follow closed-form models. 

Deep reinforcement learning (DRL) may automate such closed-loop adaptation. 
At a high level, improvement of user performance will result in larger reward for a DRL agent, which encourages it to learn the policies of adaptive time pressure that maximizes future reward, thus improving overall user performance throughout an entire cognitive task. However, DRL needs to be substantiated with a massive amount of training data. 
Even for well-defined math arithmetic tasks, it is infeasible to allow the DRL agent to interact with real users online to update its model, especially considering the slow paces of human cognitive tasks. 
In addition, whenever the hyperparameters change, the DRL model has to be retrained through a new user study. As a result, the iterative exploration of hyperparameters, which is taken for granted in machine perception tasks, becomes infeasible in such AI-mediated human cognitive tasks. 
Moreover, the initial exploration of the DRL agent training is usually random, which may jeopardize user performance.

To tackle these problems, we propose a dual-DRL agent. Specifically, a
\textit{regulation DRL agent} (Fig. \ref{teaser}) controls the existence of time pressure
feedback, while interacting with a \textit{simulation DRL agent} that mimics
real user cognition behaviors. Specifically, in the math arithmetic task,
given specific math question and time pressure feedback as input, the
simulation DRL agent could output corresponding user cognition performance,
including answering accuracy and response time. The simulation DRL agent is
pre-trained (Fig. \ref{teaser}(1)) based on an existing dataset in our prior work \cite{xu2023modeling}. It serves as virtual users to train the regulation DRL
agent (Fig. \ref{teaser}(2)), thus overcoming the aforementioned human data crunch problem. For each
math question, the regulation DRL agent decides whether time pressure 
should be delivered to virtual users according to their past performance. It
will receive more reward if virtual users' performance gets improved. This
reward design encourages the regulation DRL agent to learn potential strategies
to optimize virtual users' performance. Upon convergence, the trained model can
be used to regulate cognitive performance of real users (Fig. \ref{teaser}(3)).

With the dual-DRL agent design, the exploration capability of the DRL model can be massively enhanced as the interaction steps can be unlimited and hyperparameters can be easily iterated in the simulation environment.  
In addition, 
the random exploration of regulation DRL agent in the initial process no longer affects the real users' performance.

We have conducted an \textcolor{black}{N=80} user
study to evaluate the performance of the dual-DRL agent design.
Specifically, participants \textcolor{black}{in \textbf{RL} group} received adaptive time pressure feedback controlled by regulation DRL agent in the cognition task,
whose results were compared with \textcolor{black}{another baseline group named \textbf{Random} group}.
The final results demonstrate that the participants in
\textbf{RL} group could 
achieve better performance than \textcolor{black}{the baseline group}.
Moreover, our further analysis of \textcolor{black}{individual performance change and the} time pressure trajectory demonstrate the effectiveness of the regulation DRL agent, and explain how the regulation DRL agent could augment and improve user cognition performance. \textcolor{black}{We \textcolor{black}{have published} the software toolset and dataset\footnote{\textcolor{black}{https://github.com/songlinxu/TimeCare}} to encourage replication of this study and development of more powerful models to augment human cognition.}

\textcolor{black}{To summarize, the main contributions in this paper include:
\begin{itemize}
    \item An AI-mediated interaction mechanism to augment human cognition performance through continuous, adaptive visual feedback.
    \item A dual-DRL framework to address the data inadequacy issue in the human-in-the-loop training process. The framework could be further extended to explore various intelligent intervention/feedback to augment other human cognition tasks. 
    \item A user study of 80 participants to demonstrate the feasibility and effectiveness of our AI-mediated cognition augmentation framework.
\end{itemize}
}

\section{Related Work}
\label{sec:related}

Our work draws inspirations from and advances the knowledge in the following 4 categories of research. 

\subsection{Cognition Regulation}
Cognition performance represents observable behavior in cognitive tasks including logistic reasoning, decision making, problem solving, and memorization \cite{mayer2003causes,costa2019boostmeup}.
One of the common ways that users improve cognitive performance in their daily lives is the consumption of beverages such as coffee\cite{brice2002effects} and energy drinks\cite{childs2014influence}, as well as psychoactive drugs\cite{muller2011drugs} to increase arousal\cite{costa2019boostmeup}. However, the drinks and drugs may lead to health side effects if used excessively\cite{butt2011coffee,costa2019boostmeup}. Exercise and training\cite{steinerman2010minding} could be a safer alternative. Moreover, 
various real-time intervention systems have been developed to provide real-time feedback that regulates and improves user performance. For instance, AttentivU\cite{kosmyna2018attentivu} measures users' attention by EEG and provides feedback to increase attention, which in turn leads to better cognition performance. Recent advances in brain-computer interfaces also usher in innovations in neurofeedback, which present brain wave patterns in a proper format to help the users self-regulate their performance. For instance, online auditory neurofeedback could regulate arousal and attention in a demanding sensory-motor task\cite{faller2019regulation}. Memory could also be augmented with real-time intervention systems such as Memory Glasses\cite{corey2003memory} and SenseCam\cite{silva2013benefits}. Emotion is another important factor in cognition performance. MoodWings\cite{maclean2013moodwings} could mirror users' stress status in real-time on a wearable biofeedback device to help users calm down during stressful tasks. EmotionCheck\cite{costa2016emotioncheck} and BoostMeUp\cite{costa2019boostmeup} leveraged heart rate-like haptic feedback to regulate emotion and cognition.  

Such real-time intervention systems either present users' current emotion status to facilitate self-regulation, or provide adaptive feedback according to users' performance. The adaptive strategies are based on simple first-order heuristics. For example, AttentivU\cite{kosmyna2019attentivu} triggers vibration feedback when the EEG sensor detects a decline in user engagement. Such myopic strategies may not always increase user engagement throughout a cognitive process, and may instead act as a distraction. 
\textcolor{black}{Our work presents a closed-loop intelligent agent that could control the visual feedback adaptively according to the real-time user performance. It could learn the potential strategies to optimize the feedback over a longer horizon.}

\subsection{Reinforcement Learning for Human Behavior Regulation}

Reinforcement learning (RL) can regulate human behaviors in an interactive manner. It has been applied in diverse scenarios such as education, health, and human-machine interaction. Specifically, RL could help pacing the educational activities in the online course according to real-time learning outcomes of students, which could increase learning gains \cite{bassen2020reinforcement}. 
Liao \etal \cite{liao2020personalized} developed an RL algorithm which continuously learned to improve the treatment policy embedded in the just-in-time adaptive interventions (JITAIs) 
as context data such as step count variation collected from the user. The algorithm delivered adaptive intervention (\eg, context-tailored activity suggestion) on mobile devices to help users maintain healthy behaviors. 
In addition, an RL agent could enable a creative workflow centred on human–machine interaction, such as exploring parameter spaces in partnership with users for sound design\cite{scurto2021designing}. 

These systems verify the feasibility of leveraging RL to regulate human behaviors. However, RL training requires a large amount of data, which is challenging to obtain in human-in-the-loop training. To overcome this issue, existing work either opts for a large-scale study involving thousands of participants \cite{bassen2020reinforcement} or a highly simplified, less data-hungry algorithm \cite{liao2020personalized,scurto2021designing}. 
However, large-scale data collection is a tedious task and does not bode well with iterative model development, whereas a simplified RL algorithm may lose all the power from the deep learning wave. Even if recent human-in-the-loop RL algorithms (such as Deep COACH\cite{arumugam2019deep} and Deep Tamer\cite{warnell2018deep}) are more data-efficient than traditional RL algorithms, the hyperparameter tuning is still an iterative process that requires repetitive user studies.

\textcolor{black}{Our framework resolves this dilemma through a dual-DRL model, which first trains a simulation DRL agent to enable the generation of unlimited data, which is in turn used to train the regulation DRL agent for user intervention. This also enables hyperparamter tuning without tedious iterations of user studies.}

\subsection{Cognition Models}
Existing research in cognitive science has explored a variety of theoretical models to simulate or predict cognition performance. 
Example cognitive models include human-display interaction\cite{chen2017cognitive} and computational models such as BEAST\cite{erev2017anomalies} and Drift-Diffusion Model\cite{ratcliff2008diffusion, steyvers2019large}. Chen \etal proposed a model to capture human performance in the credit card transaction task\cite{chen2017cognitive}, by learning how eye movement patterns vary in response to data presentations. Steyvers \etal introduced \cite{steyvers2019large} a probabilistic computational model to describe different user task-switching behaviors across the lifespan\cite{steyvers2019large}. In addition to building specific models for cognition tasks such as Point-and-Click task\cite{do2021simulation,park2020intermittent}, 
RL has been used to speed up inference with user simulators in point-and-click tasks \cite{moon2022speeding}. 
Inspired by the powerful feature representation and function approximation capabilities of neural networks, deep learning-based cognitive models have been developed to 
mimic human cognition behaviors such as decision making\cite{peterson2021using, noti2016behavior, bourgin2019cognitive, plonsky2017psychological} and categorization \cite{battleday2017modeling,battleday2020capturing,singh2020end,peterson2018evaluating,battleday2021convolutional}. Notably, recurrent neural network (RNN) could simulate the tradeoff between user response time and accuracy in 
biological vision \cite{spoerer2020recurrent}. 
In addition, by integrating cognitive model priors, deep learning models could simulate human decision making behaviors more accurately\cite{bourgin2019cognitive}.

\textcolor{black}{Compared with these existing models, the cognition model in our framework leverages the recent advances of deep reinforcement learning and integrates the cognition prior model (Drift-Diffusion Model) to simulate user cognition behaviors in a more fine-grained manner. Specifically, our simulation model can mimic the progressive effect of external stimuli, instead of merely predicting the result of each cognition task.}

\begin{figure*}
\centering
\includegraphics[width=1\linewidth]{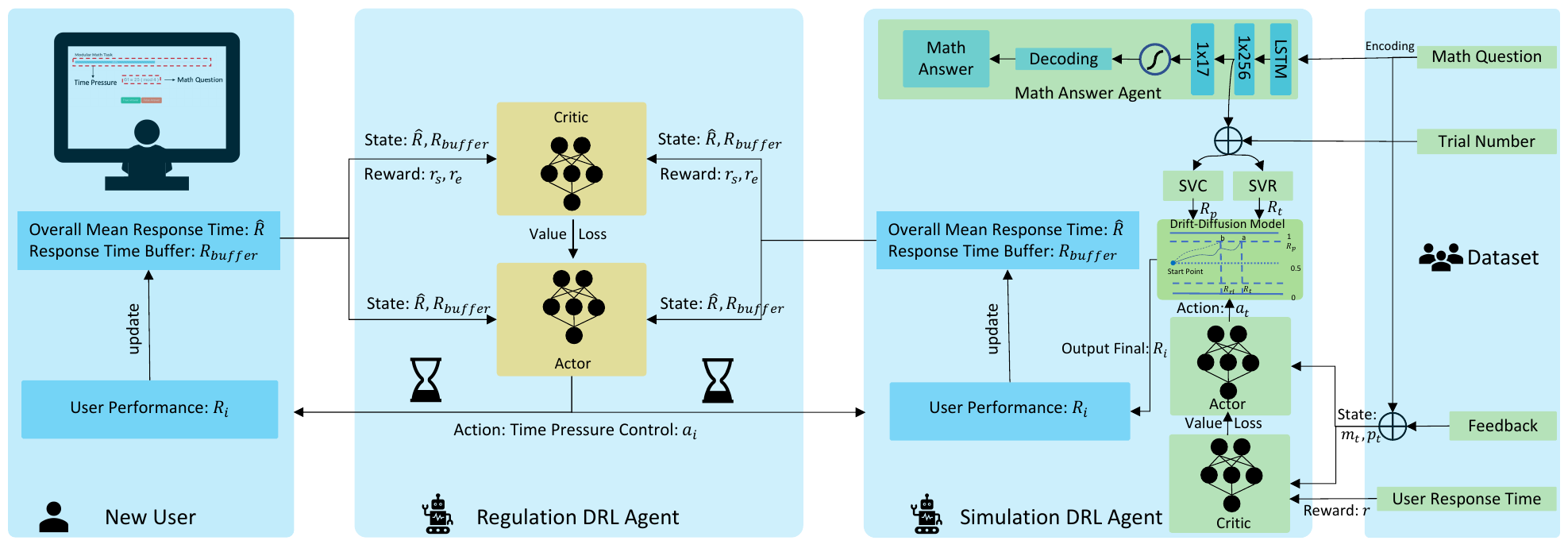}
\caption{Details of our dual DRL agents framework design. The simulation model is first pre-trained in an existing dataset. The simulation model architecture is composed of math answer agent, SVM models, DDM, and DRL loop, which aims to simulate time pressure effect on user response time in a frame-level. The final output is estimated user response time according to specific math question and time pressure feedback. The regulation DRL agent could then interact with the simulation DRL agent to explore potential time pressure control strategies to augment user cognition performance in the virtual environment. Finally, the trained regulation DRL agent is applied on real users to improve cognition performance with adaptive time pressure control. Both of the regulation DRL agent and simulation DRL agent leverage the actor-critic architecture of PPO.}
\Description{Details of our dual DRL agents framework design. The simulation model is first pre-trained in an existing dataset. The simulation model architecture is composed of math answer agent, SVM models, DDM, and DRL loop, which aims to simulate time pressure effect on user response time in a frame-level. The final output is estimated user response time according to specific math question and time pressure feedback. The regulation DRL agent could then interact with the simulation DRL agent to explore potential time pressure control strategies to augment user cognition performance in the virtual environment. Finally, the trained regulation DRL agent is applied on real users to improve cognition performance with adaptive time pressure control. Both of the regulation DRL agent and simulation DRL agent leverage the actor-critic architecture of PPO.}
\label{framework}
\end{figure*}

\subsection{Time Pressure Feedback}
Existing work has pointed out the effectiveness of time pressure in regulating human cognition \cite{moore2012time,cheng2017evaluation,slobounov2000neurophysiological}, decision making\cite{edland1993judgment}, working efficiency \cite{ban2015improving}, \etc For instance, by controlling ticking rate of a virtual clock, users' work productivity could be improved unconsciously\cite{ban2015improving}. 
Although such time pressure could improve cognitive performance by increasing focus\cite{whittaker2016don} and arousal\cite{edland1993judgment}, it may also lead to high anxiety, resulting in worse performance. In addition, different levels of time pressure may have different effects on user behaviors. For example, for risky choice behaviors, users under high time pressure tend to spend more time focusing on the negative parts, but those under low time pressure prefer to observe the positive facts\cite{zur1981effect}. Higher time pressure may also lead to excessive mental workload and deteriorated response \cite{cheng2017evaluation}. Moreover, high time pressure may increase cognition errors, as revealed in a visuomotor task study \cite{slobounov2000neurophysiological}. 

\textcolor{black}{Therefore, time pressure feedback is a mixed blessing for human cognition. Our framework is built on this insight, and leverages the dual-DRL model to capture the latent factors and tradeoffs, to elicit adaptive interventions that optimize user cognitive performance.}

\section{Cognition Task}
\label{sec:task}

We use math arithmetic task as a representative cognitive task to design and validate our framework. Math arithmetic task is widely used in evaluation of human cognition performance\cite{lin2011spatial,judd2021training,daitch2016mapping,costa2019boostmeup}. As depicted in Fig. \ref{study_setting}(a), each math question is composed of two two-digit numbers $AB$, $CD$ and one one-digit number $E$ in each trial, which is formatted as: $AB$ $\equiv$ $CD$ ( mod $E$ ). To solve this math question, participants first use $AB$ to subtract $CD$. Then they need to judge whether the result ($AB - CD$) is divisible by $E$. We developed a web application to present the math problems to users one after another. A user simply selects the "True" or "False" button on the screen in response to each problem. We use accuracy and response time as the cognitive performance metrics.

\section{Time Pressure Design}
\label{time pressure}

\begin{figure*}
\centering
\includegraphics[width=1\linewidth]{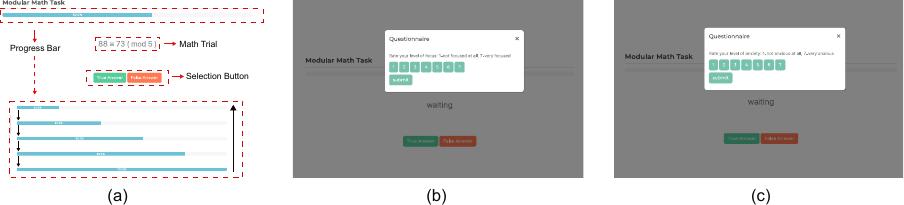}
\caption{(a). User interface in our study setting. The math question is shown in the center of the screen. The time pressure feedback (progress bar) is delivered on top of the math question, which increases one unit per second and will reset after 5 seconds. (b,c). User interface where participants are asked to rate their current attention (b)/ anxiety (c) status.}
\Description{(a). User interface in our study setting. The math question is shown in the center of the screen. The time pressure feedback (progress bar) is delivered on top of the math question, which increases one unit per second and will reset after 5 seconds. (b,c). User interface where participants are asked to rate their current attention (b)/ anxiety (c) status.}
\label{study_setting}
\end{figure*}

We adopt time pressure as the feedback modality in our dual-DRL framework, as its effectiveness in regulating user cognition has been well established, particularly for tasks such as attention  \cite{cheng2017evaluation,slobounov2000neurophysiological}, decision making\cite{edland1993judgment} and productivity\cite{ban2015improving}. Time pressure could be delivered in different forms, \eg, haptic, auditory, and visual feedback. Here we adopt visual feedback as it can be delivered to users more quickly and efficiently.
To convey a sense of urgency without aggravating the cognitive workload, and in particular to keep users from focusing too much on the exact time, 
we present to users a simple progress bar instead of a clock as the visual feedback, along with the math problem. 
The progress bar adds one unit per second (Fig. \ref{study_setting}(a)). It resets after 5 seconds and starts increment again afterwards. In this way, users will only receive a sense of urgency due to time passage, but will not focus on the exact time.

\section{Dual-DRL Agent Framework}
\label{sec:framework}

The overarching goal of our framework is to train a regulation DRL agent that controls the existence of time pressure feedback during each \textcolor{black}{trial, \ie, each process where a math question is presented and then answered by the user.} 
This regulation DRL agent is first trained in a simulation environment, which mimics cognition behaviors of real users. Then the trained regulation DRL agent is directly applied to regulate real users in real-time. 

In our framework, the simulation environment is actually another DRL agent (\ie, the \textit{simulation DRL agent}), which learns to mimic real user behaviors based on an existing dataset. \textit{Given a specific math question, trial number and feedback pattern as input, the simulation DRL agent will output the corresponding answer and response time, just like what real users will do in the cognition task}. In this way, the regulation DRL agent could always interact with the simulation DRL agent to explore potential time pressure control strategies to regulate user cognition performance. The online training data is unlimited to ensure the convergence of regulation DRL agent.
We could also fine-tune the hyperparameters of regulation DRL agent and apply them into the new round of interaction with the simulation DRL agent. In addition, the random exploration of regulation DRL agent in the initial training stage will only happen in the simulation environment instead of real users. Upon convergence, the trained regulation DRL agent could be applied into real-world scenarios. 
In what follows, we elaborate on the architecture details of both  simulation DRL agent and regulation DRL agent, respectively.

\subsection{Simulation DRL Agent}
\label{sec: sub: simulation drl}
The model architecture of the simulation DRL agent follows the design of our recent work \cite{xu2023modeling}. We thus only briefly summarize the design to avoid redundancy. \textcolor{black}{Before modeling the effect of time pressure, we first need to model users' basic cognition performance without time pressure, which is achieved by an LSTM (Long Short-Term Memory)-based math answer agent and an SVM (Support Vector Machine) model (Fig.~\ref{framework}). The math answer agent is trained to answer math questions correctly instead of predicting user performance. By doing so, the math answer agent could extract potential features, in the form of intermediate output (Fig. \ref{framework}), which reflect user performance change across different math questions. These features are then fed into an SVM model to predict a user's basic performance (both accuracy and response time). }

Then, in order to simulate the effect of time pressure on user performance in a fine-grained manner, we segment cognition process into steps. For each step,
the corresponding image frame of time pressure feedback as well as encoded math questions are fed into a DRL loop to simulate the effect of time pressure on user performance. The final output is a prediction of user response time. 
\textcolor{black}{Note that one frame represents the corresponding time pressure visual image during one step}. \textcolor{black}{Each math trial is segmented into steps with frequency $f=5$ Hz. For instance, if user response time for one trial is 5 seconds, then the math trial comprises 25 steps. \textcolor{black}{The selection of different frequency values has been discussed in \cite{xu2023modeling}}. Such segmentation affords a more fine-grained model to simulate the effect of time pressure. } 

Note that here the DRL agent only simulates user response time under time pressure; whereas the response accuracy is predicted directly from the SVM models. The reason is that users' speed-accuracy performance is usually biased by the instruction \cite{park2020intermittent,zhai2004speed}. In our experiments, \textit{the participants are asked to take accuracy as priority and then answer questions as soon as possible. As a result, user accuracy will not be affected by time pressure significantly}. Therefore, we only use the DRL agent to simulate the effect of time pressure on users' response time instead of accuracy.

\subsubsection{Math Answer Agent}
The math answer agent is a sequence-to-sequence model based on LSTM, which is trained to predict math question answers. 
We first use sequence encoding to encode each math question string into a sequence vector. The sequence vector is then fed into LSTM layer (256 neurons in the hidden unit), which is then connected with 17 neurons with the softmax activation function. Finally, the neuron with the highest probability is the final output answer. We use Keras\cite{chollet2015keras} to implement the math answer agent model (loss function: categorical cross entropy, optimizer: Adam, learning rate: 0.001).

\subsubsection{SVM Model: Predicting Baseline Performance}
The SVM models take the features captured by the math answer agent and trial number as input, and predict user's choice and response time without time pressure. 
We use an SVM classifier (SVC) and regressor (SVR) to predict user choice $R_c$ and response time $R_t$, respectively. The SVM models are implemented with scikit-learn\cite{scikit-learn}, where we use default regularization parameter, kernel, and other parameters for both SVC and SVR. 
The SVC not only predicts user choice but also outputs the probability $R_p$ for each possible choice.

\subsubsection{Drift-Diffusion Model and DRL Agent}
Our simulation DRL agent relies on a widely-used cognition model, \ie Drift-Diffusion Model (DDM) \cite{fudenberg2020testing} to mimic user cognition performance. DDM assumes that users make selection by accumulating evidence for each choice and the final selection is made when the evidence accumulator passes a boundary threshold \cite{fudenberg2020testing}. 
Here we incorporate the SVM model prediction results into DDM. Specifically, we use the output probability of SVC as the accumulated evidence, whose start point is 0.5. The boundary threshold is $R_p$, which is the probability when SVC makes the predictions. When there is no time pressure, the evidence accumulator will increase from start point to boundary threshold ideally. However, when time pressure is delivered to users, the trajectory of evidence accumulator will change and the evidence accumulator may achieve the boundary threshold earlier or later than before, resulting in the new response time of users. 

We incorporate a DRL agent to simulate this effect of time pressure on evidence accumulation process. Specifically, we segment the initial evidence accumulation process into steps under specific frequency $f=5$ Hz. In each step, we leverage the DRL agent to simulate the effect of corresponding frame of time pressure visual feedback on evidence accumulation process. As a result, the evidence accumulator may achieve boundary threshold earlier or later than before, which will be the final estimated user response time.

\subsubsection{Action Space}
For each step $t$ in the evidence accumulation process, the action ($a_t$) of the simulation DRL agent is a continuous numeric value normalized from $-1$ to $1$, 
\ie $a_t \in [-1,1]$. 
$a_t = 0$ means that current time pressure frame has no effect on evidence accumulator in DDM. $a_t \in [-1,0)$ means the current time pressure frame introduces negative change on the evidence accumulator, and positive otherwise. 

\subsubsection{Observation Space}
The observation space (State: $S_t = (m_t, p_t)$) of the simulation DRL agent is composed of both math question information and dynamic time pressure visual stimuli. Similar to the math answer agent, for each math trial, the math question information is encoded as a sequence vector ($m_t$) in the observation space. In accordance to the segmentation of evidence accumulation process, the dynamic time pressure visual stimuli is also segmented into frames just like what users will receive in the study. Given frame rate $f$, for each frame $t$, we could obtain the specific image $p_t$ of time pressure visual stimuli for input in the observation space, which will be extracted into specific features from the default CNN feature extractor in 
Stable Baselines3 \cite{stable-baselines3}.

\subsubsection{Terminal State}
The terminal state is achieved if the evidence accumulator achieves boundary threshold ($R_p$) or the DRL agent achieves maximum steps in one \textcolor{black}{episode}. \textcolor{black}{Here each episode is composed of several steps. When the DRL agent achieves the terminal state, one episode ends and a new episode will start. In this simulation DRL agent, one episode is the same as one math trial}. 
We set the maximum response time $RT_{max}$ to be 10 seconds. So the maximum step number $S_N = RT_{max} \times f=10 \times 5=50$ steps. If the evidence accumulator achieves boundary threshold $R_p$ when DRL agent takes $S_n$ steps ($S_n < S_N$), then the new predicted response time is $R_{rl} = S_n/f$.

\subsubsection{Reward Function Design}
For each step during an episode, the simulation DRL agent only gets reward in the terminal state. The reward is 0 for other situations. 
The reward function aims to encourage the simulation DRL agent to mimic real user response time better than the SVM models. Therefore, the reward function in terminal state can be denoted as: 

\begin{equation}
    r = \begin{cases}
    |E_{rl}-E_{svm}|/E_{svm} + P^*, \quad & E_{rl}<E_{svm} \\
    0, \quad & E_{rl} \geq E_{svm}
    \end{cases}
\end{equation}

\noindent where $E_{rl}$ and $E_{svm}$ are estimated error rate of simulation DRL predicted response time ($R_{rl}$) and SVM predicted response time ($R_{svm} = R_t$) compared with real response time ($R_u$) of users respectively, \ie, $E_{rl} = |R_{rl}-R_u|/R_u$, $E_{svm} = |R_{svm}-R_u|/R_u$. $P^*$ is the penalty caused by the terminal state. If the simulation DRL agent's step number exceeds the maximum step threshold $S_N$, then $P^* = -1$. Otherwise, $P^* = 0$.

\subsubsection{Policy Optimization Algorithm}
We use Proximal Policy Optimization (PPO)\cite{Schulman2017ProximalPO} as the learning algorithm and multilayer perceptron (MLP) to be the policy for agent training. PPO is an actor-critic algorithm that is based on a policy gradient method. The actor network aims to maximize the loss function below:

\begin{equation}
    L(\theta) = \mathbb{E}_t[\min (\frac{\pi_\theta(a|s)}{\pi_{\theta_{old}}(a|s)} A(s,a), clip(\frac{\pi_\theta(a|s)}{\pi_{\theta_{old}}(a|s)},1-\epsilon,1+\epsilon)A(s,a))]
\end{equation}

\noindent where $\pi_\theta(a|s)$ is the probability of choosing action $a$ given state $s$ and weights $\theta$ of actor network. 
The clipping method\cite{Schulman2017ProximalPO} is used to avoid large gradient updates that may make the training process unstable. $A(s,a)$ calculates relative benefit of choosing action $a$, \ie

\begin{equation}
    A(s_t,a_t) = R(s_t,a_t) + \gamma V_\phi(s_{t+1}) - V_\phi(s_t)
\end{equation}

\noindent where $\gamma$ serves as the discounting factor to encourage the agent to accumulate rewards faster and $V_\phi(s_t)$ is from the critic network given weights $\phi$. We refer readers to  \cite{Schulman2017ProximalPO} for more details of PPO.

\subsubsection{Agent Training}

The simulation DRL agent is trained in an existing dataset collected in our previous work \cite{xu2023modeling}, which contains 25,000 math problem trials from real users. 
We have trained different simulation DRL agents that contain not only a general model from the overall dataset but also personalized models that are trained from specific users respectively. 
In this paper, we mainly aim to provide a simulation environment that could represent general cognition behaviors of real users to interact with the regulation DRL agent. Therefore, we adopt the general model instead of personalized models. 
The predicted response time in the general model of simulation DRL agent could achieve 0.3022 of MAPE \textcolor{black}{(Mean Average Percentage Error)} in the testing set and it could also capture the general trend of user response time changes, \textcolor{black}{which is depicted in\cite{xu2023modeling}}. \textcolor{black}{Here MAPE represents the average of the absolute percentage errors of predictions. Given a prediction array $\hat{y}_i (0 < i \leq n)$ and a label array $y_i (0 < i \leq n)$, MAPE is calculated as $MAPE = \frac{1}{n} \sum_{i=1}^{n} |\frac{\hat{y}_i-y_i}{y_i}| $.   }

\subsection{Regulation DRL Agent}

\textcolor{black}{Unlike the simulation DRL agent that aims to mimic user cognition behaviors, the regulation DRL agent aims to control the existence of time pressure to regulate user cognition performance according to users' real-time cognition results. Users (either a real user or virtual one from the simulation DRL agent) will serve as the environment that interacts with the regulation DRL agent. For each math trial, users produce certain answers which, along with response time, will be passed to the observation space of the regulation DRL agent. According to the past user cognition results, the regulation DRL agent will decide whether the time pressure feedback should be delivered to users in order to maximize the overall user cognition performance. We now provide more details below.  }

\begin{figure*}
\centering
\includegraphics[width=1\linewidth]{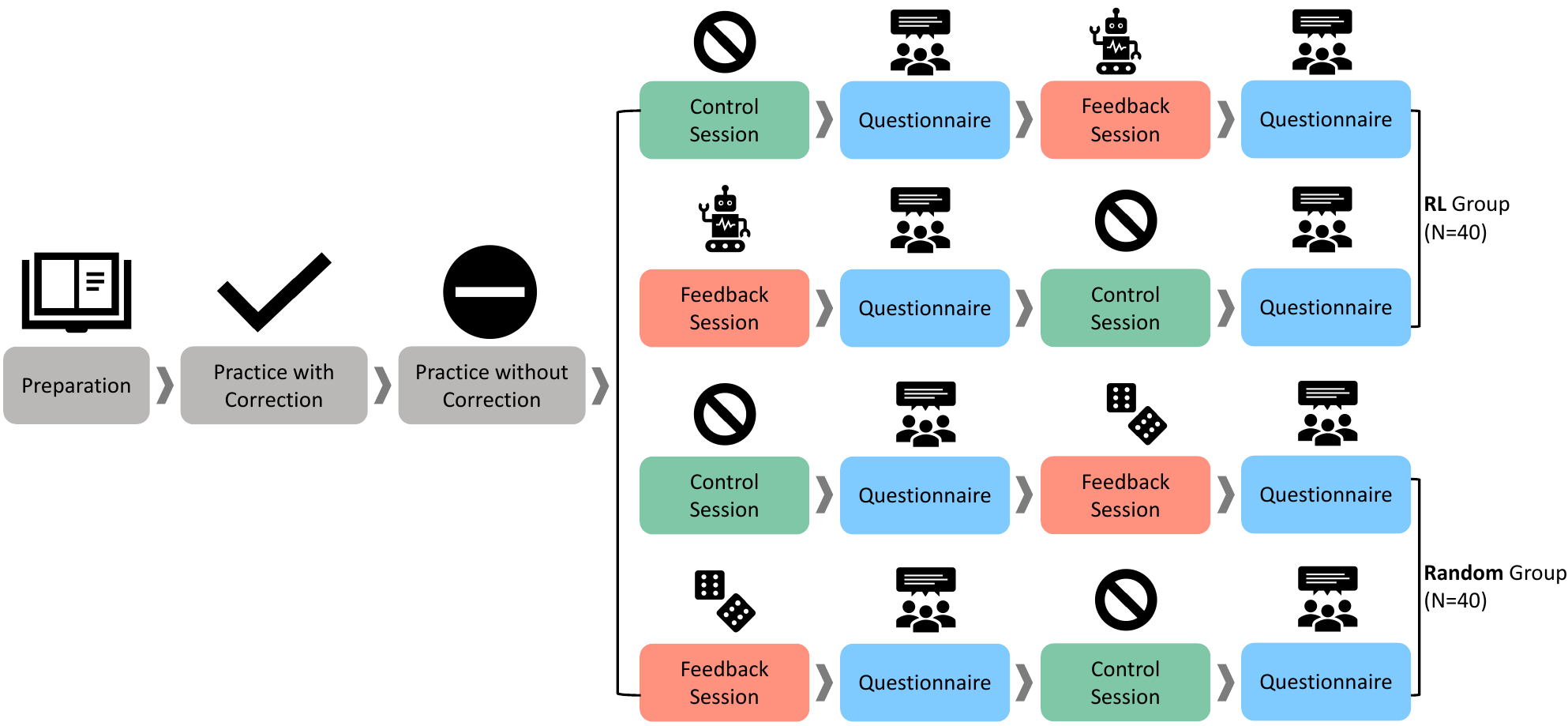}
\caption{\textcolor{black}{An illustration for our study design.}}
\Description{An illustration for our study design.}
\label{study_design}
\end{figure*}

\subsubsection{Action Space}
The regulation DRL agent controls the existence of time pressure feedback in each trial to regulate user performance. Therefore, for each step $i$, the regulation DRL agent takes binary action ($a_i \in \{0,1\}$) during each trial. If $a_i = 1$, the time pressure feedback will be delivered to users in the current trial, and no feedback otherwise. 
\textcolor{black}{Note that each step here corresponds to one math trial, which is different from the definition of step in the simulation DRL agent. The reason is that the simulation DRL agent aims to simulate the progressive effect of time pressure on user cognition performance by segmenting user cognition process of each trial into steps. Therefore, \textcolor{black}{each step of the simulation DRL agent corresponds to each frame of time pressure feedback}. However, the regulation DRL agent aims to control the existence of time pressure during each trial to regulate user performance. Therefore, each step of the regulation DRL agent is one math trial.
}

\subsubsection{Observation Space}
As depicted in Sec. \ref{sec: sub: simulation drl}, our study settings require participants to take accuracy as priority. As a result, user accuracy will not be significantly affected by time pressure and response time is mainly used to reflect user performance. Therefore, the regulation DRL agent will mainly care about user response time. Specifically, the state of regulation DRL agent is denoted as $S_i = (\hat{R}, R_{buffer})$, where $\hat{R}$ is the overall average response time of users which will update continuously as the trials proceed. Additionally, $R_{buffer}$ represents user response time of the most recent 10 trials, \ie, $R_{buffer} = [R_1, R_2, \cdots, R_{10}]$, which is also updated continuously. This design ensures the regulation DRL agent could be both far-sighted to optimize the long-term average of user response time, and sensitive enough to boost the immediate response.

\subsubsection{Terminal State}
Terminal state happens when the regulation DRL agent achieves maximum steps in one episode. In order to simulate real user studies, we set the maximum step number $N$ to be the same as the number of all trials for each participant in our existing dataset. When the regulation DRL agent achieves terminal state, a new \textcolor{black}{episode} will begin, just like a new user study for one participant will begin. \textcolor{black}{Note that the episode here consists of several math trials, unlike that in the simulation DRL agent. In the regulation DRL agent, each episode could be viewed as each user study for one participant.}

\subsubsection{Reward Function Design}
The reward is composed of two parts: step reward $r_s$ and end reward $r_e$. Step reward represents the reward that the agent will obtain during each step, \ie each math trial. End reward stands for the reward that the agent will get in terminal state, \ie one episode is finished. The reward function aims to encourage regulation DRL agent to improve user cognition performance (response time). Therefore, we first set user initial response time to be $R_{init}$ when there is no time pressure, which is actually obtained from existing dataset from our previous work \cite{xu2023modeling}. In fact, in our investigation of model design, we found that the value of $R_{init}$ will not significantly affect performance of regulation DRL agent. Because the agent will mainly care about relative change of user response time, instead of absolute value. We also denote user response time at step $i$ to be $Ru_i$. Then we have:

\begin{equation}
    r_s = R_{init} - Ru_i
\end{equation}

\begin{equation}
    r_e = \begin{cases}
    (\Delta_{Ru}-\Delta_{R^*})/\Delta_{R^*}, \quad & i = N \\
    0, \quad & i < N
    \end{cases}
\end{equation}

\begin{table*}[]
\caption{\textcolor{black}{Calculation for Relative and Absolute Evaluation Metrics. $c$: \textbf{Control} session, $f$: \textbf{Feedback} session.}}
\label{tab:metric_calculate}
\begin{tabular}{l|l|l}
\hline
Type & Metric  & Formula \\ \hline
\multirow{4}{*}{Absolute}   & accuracy & $\Delta_{absolute\ accuracy} = Accuracy_{f} - Accuracy_{c}$\\ \cline{2-3} 
                                         & response time & $\Delta_{absolute\ response\ time} = Response\ Time_{f} - Response\ Time_{c}$\\ \cline{2-3} 
                                         & attention     & $\Delta_{absolute\ attention} = Attention_{f} - Attention_{c}$\\ \cline{2-3} 
                                         & anxiety       & $\Delta_{absolute\ anxiety} = Anxiety_{f} - Anxiety_{c}$\\ \hline
\multirow{4}{*}{Relative}   & accuracy &$\Delta_{relative\ accuracy} = (Accuracy_{f} - Accuracy_{c})/Accuracy_{c}$\\ \cline{2-3} 
                                         & response time & $\Delta_{relative\ response\ time} = (Response\ Time_{f} - Response\ Time_{c})/Response\ Time_{c}$ \\ \cline{2-3} 
                                         & attention     & $\Delta_{relative\ attention} = (Attention_{f} - Attention_{c})/Attention_{c}$ \\ \cline{2-3} 
                                         & anxiety       & $\Delta_{relative\ anxiety} = (Anxiety_{f} - Anxiety_{c})/Anxiety_{c}$ \\ \hline
\end{tabular}
\end{table*}

\noindent where $\Delta_{Ru}$ is the overall relative reduction of user response time in total, \ie $\Delta_{Ru} = (R_{init}-\hat{Ru}) /R_{init}$. Here $\hat{Ru}$ is the average response time in all trials that have been finished by users, which will be updated during each step. $\Delta_{R^*}$ is the relative reduction target that we set for the agent. In other words, the agent will get more end reward if overall relative reduction of user response time $\Delta_{Ru}$ is larger than the target reduction $\Delta_{R^*}$. We set the target reduction $\Delta_{R^*}$ to be 0.1054, which is obtained from the relative response time reduction of the best group (\textbf{Random} group, will be introduced in Sec. \ref{sec: user study new}) in our existing dataset. 

In summary, the end reward aims to encourage the agent to achieve better improvement of user cognition performance than that of the best baseline group in our existing dataset. The best baseline group utilizes a random strategy to control time pressure feedback, which will be described in Sec.~\ref{sec: user study new}.

\subsubsection{Policy Optimization Algorithm}
Similar with the simulation DRL agent, we use Proximal Policy Optimization (PPO)\cite{Schulman2017ProximalPO} as the learning algorithm and multilayer perceptron (MLP) to be the policy for agent training. Both the simulation DRL agent and the regulation DRL agent are implemented with PyTorch\cite{NEURIPS2019_9015}, Stable Baselines3\cite{stable-baselines3}, and Gym\cite{brockman2016openai}.

\subsubsection{Agent Training}
As depicted before, we first let the regulation DRL agent interact with the simulation DRL agent all the time until the regulation DRL agent finally converges at approximately 50000 steps. We also test the trained regulation DRL agent in the simulation environment. The result shows that the improvement of user cognition performance can indeed achieve our target. A real user study that demonstrates the superiority of the regulation DRL agent and specific description of other baseline groups will be introduced in Sec. \ref{sec: user study new}.

\section{User Study}
\label{sec: user study new}

To evaluate whether the trained regulation DRL agent could effectively control the time pressure feedback and regulate human cognition, a user study was conducted to examine how users' performance is improved under this adaptive feedback.

\subsection{Participants}

We recruited \textcolor{black}{80} participants (\textcolor{black}{age 20.13 ± 1.86 y (mean ± SD); 36 female}) from the campus of a large US public university. Participants came from a variety of majors including engineering, computer science, mathematics, psychology, and so on. One participant's data was removed due to technical issues in the study. So the effective participant number is \textcolor{black}{79}. \textcolor{black}{We got approval from the local IRB office prior to the study.} 

\subsection{Apparatus and Task}

All participants were asked to finish the math arithmetic task, which was depicted in Fig.~\ref{study_setting}(a) and Sec. \ref{sec:task}. We developed a web application platform, where participants directly logged into the study through a unique user code. The math question during each trial as well as time pressure visual feedback was presented to participants on the screen like Fig. \ref{study_setting}(a). For each math trial, participants were required to make a binary selection, as depicted in Sec.~\ref{sec:task}. The user response time during each trial was also recorded by the system. Participants were instructed to make the selection immediately after they had the answer. Before the study, participants were asked to prepare themselves and avoid potential interruptions during the study.

\subsection{Study Procedure}

All participants were first instructed to read and sign the consent form of the study. \textcolor{black}{The experimenter answered all the participants' questions to make sure they understood the consent form.} The experimenter then introduced the study procedure and math arithmetic tasks. After that, the participants would first run a \textcolor{black}{first practice session} (10 math trials) to get familiar with the math task where no time pressure was provided, and the system would also tell them whether their answers were correct or not. Participants were allowed to extend this practice session if they needed to further familiarize themselves with the task. Then the participants would have a \textcolor{black}{short rest for 10 seconds}, after which they continued the \textcolor{black}{second practice session (10 trials)}, where no time pressure was shown but the system would not tell them whether their answers were correct or not. Another 1 minute of rest followed the second practice session. Finally, participants would start the last session \textcolor{black}{(formal session: 200 trials), which was composed of two test sessions (100 trials for each). One test session was the \textbf{Control} session, where no feedback would happen. Another test session was the \textbf{Feedback} session, where feedback would be controlled by a specific strategy. The order of the two sessions were counter-balanced, whose details will be described later. After each test session, participants would fill in a short questionnaire to rate their attention and anxiety level in a 7-point likert scale (Fig.~\ref{study_setting}(b,c)) and have a 2-min rest.} It took each participant about \textcolor{black}{20-30} minutes for the whole study. After the study, we also had an informal interview to ask the participants about their experience during the study.

\begin{figure*}
\centering
\includegraphics[width=1\linewidth]{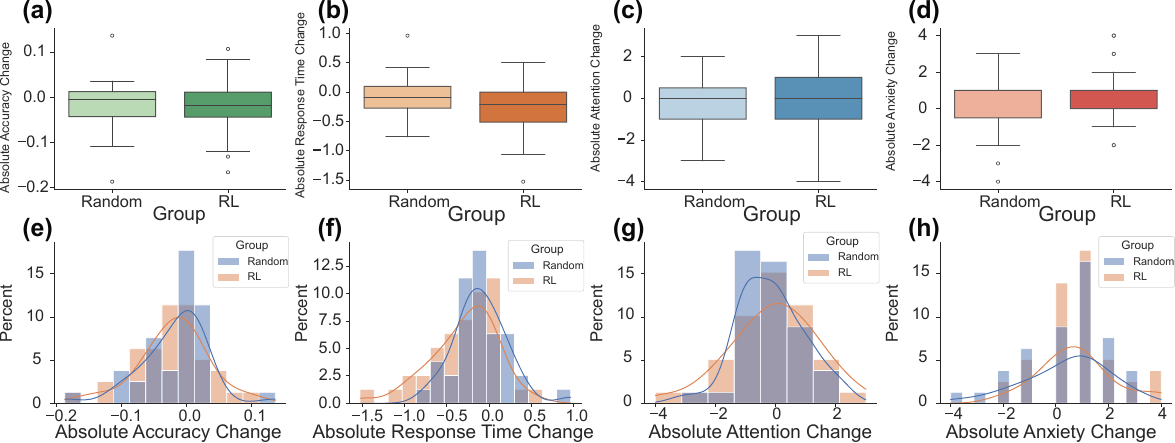}
\caption{\textcolor{black}{Boxplot (first row) and distribution plot (second row) of \textbf{absolute} delta change of accuracy (a,e), response time (b,f), attention (c,g), anxiety (d,h) from \textbf{Control} session to \textbf{Feedback} session in two groups. }}
\Description{Boxplot (first row) and distribution plot (second row) of \textbf{absolute} delta change of accuracy (a,e), response time (b,f), attention (c,g), anxiety (d,h) from \textbf{Control} session to \textbf{Feedback} session in two groups. }
\label{u2_abs}
\end{figure*}

\subsection{Experimental Design}
\label{sec: experiment design}

\textcolor{black}{We utilized a between-subject design similar to BoostMeUp \cite{costa2019boostmeup}. 80 participants were randomly and uniformly divided into two groups: \textbf{RL} and \textbf{Random} group. Each group of participants would finish two main test sessions: \textbf{Control} session and \textbf{Feedback} session. The order of the two sessions in the two groups was counter-balanced. As depicted in Fig. \ref{study_design}, for each group, half of the participants \textcolor{black}{(N=20)} first experienced the \textbf{Control} session and another half \textcolor{black}{(N=20)} first experienced the \textbf{Feedback} session. 
By doing so, we could compare user performance improvement from different feedback control strategies using both relative and absolute metrics.
}

\textcolor{black}{For \textbf{RL} group, each participant received adaptive time pressure feedback, controlled by our regulation DRL agent according to their past performance in \textbf{Feedback} session. No time pressure feedback was provided in \textbf{Control} session. }

\textcolor{black}{For \textbf{Random} group, each participant received random feedback in \textbf{Feedback} session, meaning that there was 50\% chance that time pressure would happen during each math trial. No time pressure was provided in \textbf{Control} session. }

\textcolor{black}{The study procedure was the same for both \textbf{RL} and \textbf{Random} group. Feedback control strategy in \textbf{Feedback} session was the only difference between the two groups.}

\textcolor{black}{This \textbf{Random} group serves as baseline for comparison with the \textbf{RL} group. The reason is that our previous work \cite{xu2023modeling} has shown that random feedback leads to better user performance compared with none feedback, static feedback, or even linearly-controlled closed-loop feedback\footnote{\textcolor{black}{None Feedback: no feedback is provided. Static Feedback: time pressure feedback is provided during each trial. Linear-Controlled Closed-Loop Feedback: the existence of time pressure feedback will be controlled in a linear closed-loop manner according to users' real-time performance, \ie, time pressure is cancelled when user performance drops and activated when user performance does not drop. More details in \cite{xu2023modeling}.}}. The reason is that users may just get used to none/static/linear-controlled feedback over time, and may not be regulated effectively. However, users could not anticipate when the feedback may happen in \textbf{Random} group.}

\textcolor{black}{There were 100 math trials in both \textbf{Control} and \textbf{Feedback} session, respectively. \textcolor{black}{Too few trials may not provide enough evidence for accurate statistics, while too many trials may make} participants too tired and bring irrelevant artifacts to the results. 100 trials are the appropriate choice according to our pilot study.
After each test session, participants would fill in a short questionnaire to rate their attention and anxiety level in a 7-point likert scale (Fig. \ref{study_setting}(b,c)) and have a 2-min rest. The 2-min rest could make sure participants refresh themselves which is verified in our pilot study to avoid effect across two sessions. Before the two test sessions, there were two practice sessions. The first practice session (10 trials) would tell participants whether their answers were correct, \textcolor{black}{which aimed to help them get familiar with our task}. The second practice session (10 trials) would not tell participants the correctness of their answers, \textcolor{black}{which aimed to help them get familiar with the situation where they would not receive notification about their correctness. Because the two formal test sessions would not tell them their correctness as well}. 10 trials are enough to let participants get familiar with our tasks according to our pilot study. \textcolor{black}{Too few trials may not make sure that users could understand our task well, while too many trials may let users tired and bring irrelevant artifacts to other sessions}. The whole study design and procedure is depicted in Fig.~\ref{study_design}. 
}

\begin{figure*}
\centering
\includegraphics[width=0.99\linewidth]{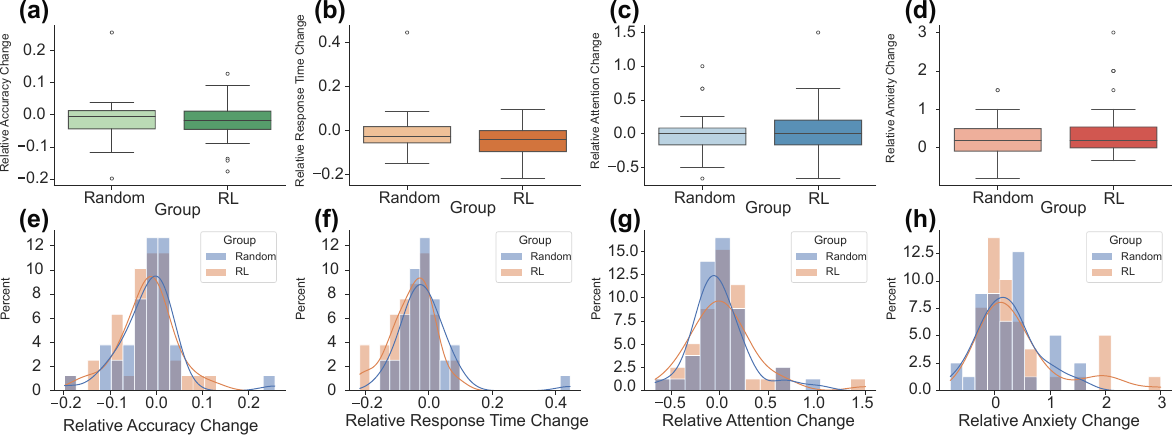}
\caption{\textcolor{black}{Boxplot (first row) and distribution plot (second row) of \textbf{relative} delta change of accuracy (a,e), response time (b,f), attention (c,g), anxiety (d,h) from \textbf{Control} session to \textbf{Feedback} session in two groups. }}
\Description{Boxplot (first row) and distribution plot (second row) of \textbf{relative} delta change of accuracy (a,e), response time (b,f), attention (c,g), anxiety (d,h) from \textbf{Control} session to \textbf{Feedback} session in two groups. }
\label{u2_rela}
\end{figure*}

Recall that we encouraged participants to prioritize accuracy over response time.  In order to encourage participants to try their best to perform well in the task, the compensation also varied according to their performance. Specifically, we informed the participants that we would first compare the average accuracy of them and then break even by comparing the average response time afterwards.
\textcolor{black}{The top 1-3, 4-9, 10-15, and 16-30 participants would receive $\$ 70$, $\$ 50$, $\$ 35$, $\$ 30$, respectively. Other participants would receive $\$ 20$.} Moreover, in order to encourage participants to take accuracy as the priority, for each wrong answer they might made, they would lose $\$ 1$, until their final reward achieved $\$ 10$. Therefore, all participants would receive at least $\$ 10$. But we used more rewards to encourage participants to follow our instructions and take the study seriously and carefully.

\subsection{Experimental Metrics}

\textcolor{black}{Our study revolves 4 evaluation metrics: accuracy, response time, attention score, and anxiety score, to capture the change of user cognition performance and user emotion status. We calculate the average accuracy and average response time for both \textbf{Control} session and \textbf{Feedback} session. Then we get the delta change of accuracy ($\Delta_{accuracy} = accuracy_{feedback} - accuracy_{control}$) and response time ($\Delta_{response\ time} = response\ time_{feedback} - response\ time_{control}$) from \textbf{Control} session to \textbf{Feedback} session, which are compared in both \textbf{RL} and \textbf{Random} groups. Similar calculation method works for attention and anxiety score.}

\textcolor{black}{We notice that \textit{different participants' math problem solving abilities may vary}. Therefore, we decide to compare both of the \textbf{relative} and \textbf{absolute} delta change of user performance and emotion status in order to eliminate users' individual differences and elucidate the impact of time pressure across different groups. The absolute delta is just calculated as above. The relative delta change is calculated using the absolute change divided by the value in the \textbf{Control} session. Finally, both \textbf{relative}/\textbf{absolute} change of accuracy, response time, attention score and anxiety score are depicted in Table~\ref{tab:metric_calculate}.}

\subsection{Evaluation Results}

\textcolor{black}{We performed a two-way between-subject ANOVA to analyze the results and compare the effect of Groups (\textbf{RL}, \textbf{Random} group) and Orders (Order 1: \textbf{Control} session first, Order 2: \textbf{Feedback} session first).
We removed abnormal data samples whose response time was smaller than 0.8 seconds or larger than 10 seconds, which might appear when participants triggered the keyboard accidentally or got absent-minded in the study.
}

\subsubsection{Response Time}

\textcolor{black}{For \textbf{absolute} response time delta change, no significant effect was found for Order ($F_{1,75} = 2.476, P = 0.12 > 0.05$). However, significant effect was found for Group ($F_{1,75} = 6.309, P = 0.014 < 0.05$), \ie, \textbf{RL} group (mean $\pm$ SD: $ -0.297 \pm 0.400 $) v.s. \textbf{Random} group (mean $\pm$ SD: $ -0.094 \pm 0.320 $). No significant interaction was found between Group and Order ($F_{1,75} = 0.407, P = 0.525 > 0.05$). Fig.~\ref{u2_abs}(b) also corroborates the conclusions above, where we find that participants in \textbf{RL} group have larger absolute response time reduction from \textbf{Control} to \textbf{Feedback} session, compared with participants in \textbf{Random} group. The distribution of absolute response time delta change in both groups is also depicted in Fig.~\ref{u2_abs}(f), where we see that more data samples distribute at the negative response time absolute change (reduction) in \textbf{RL} group compared with \textbf{Random} group, indicating that the regulation DRL agent could decrease user response time better than random feedback in this absolute evaluation metric.}

\textcolor{black}{For \textbf{relative} response time change, no significant effect was found for Order ($F_{1,75} = 0.678, P = 0.413 > 0.05$). However, significant effect was found for Group ($F_{1,75} = 5.253, P = 0.025 < 0.05$), \ie, \textbf{RL} group (mean $\pm$ SD: $ -0.053 \pm 0.069 $) v.s. \textbf{Random} group (mean $\pm$ SD: $ -0.011 \pm 0.092 $). No significant interaction was found between Group and Order ($F_{1,75} = 0.506, P = 0.479 > 0.05$). Similar with \textbf{absolute} results, we find larger \textbf{relative} response time reduction in \textbf{RL} group compared with \textbf{Random} group in the boxplot of Fig.~\ref{u2_rela}(b) and distribution plot of Fig. \ref{u2_rela}(f).
}

\textcolor{black}{To summarize, both \textbf{absolute} and \textbf{relative} evaluation metrics of response time delta change support the conclusion that participants in \textbf{RL} group have better cognition performance than \textbf{Random} group, demonstrating the effectiveness of our regulation DRL agent in augmenting user cognition performance.}

\begin{figure*}
\centering
\includegraphics[width=1\linewidth]{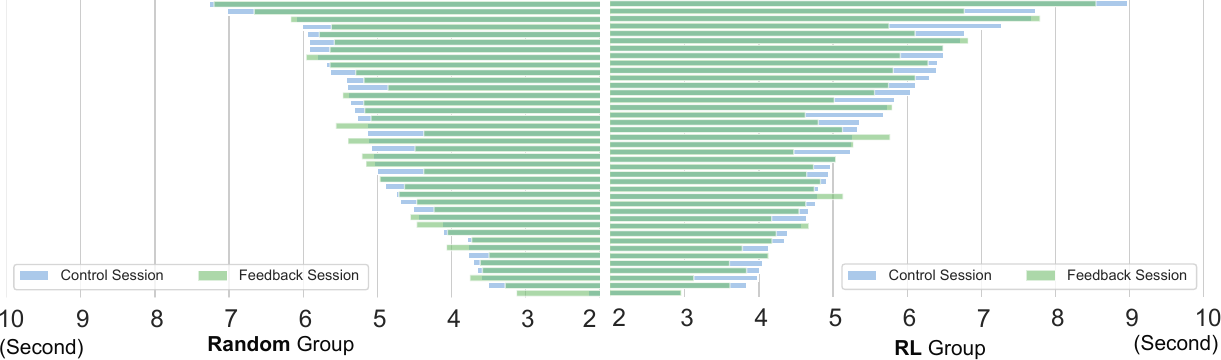}
\caption{\textcolor{black}{Bar plot for individual \textbf{absolute} response time in \textbf{Random} group (left) and \textbf{RL} group (right). Each bar represents one participant in the corresponding group. The blue bar and green bar represent individual response time in \textbf{Control} session and \textbf{Feedback} session, respectively. The bar plot is sorted according to the absolute average response time in \textbf{Control} session.}}
\Description{Bar plot for individual \textbf{absolute} response time in \textbf{Random} group (left) and \textbf{RL} group (right). Each bar represents one participant in the corresponding group. The blue bar and green bar represent individual response time in \textbf{Control} session and \textbf{Feedback} session, respectively. The bar plot is sorted according to the absolute average response time in \textbf{Control} session.}
\label{resptime_individual}
\end{figure*}

\subsubsection{Accuracy}
\textcolor{black}{In terms of \textbf{absolute} accuracy delta change, no significant effect was found for Order ($F_{1,75} = 1.343, P = 0.25 > 0.05$), or Group ($F_{1,75} = 0.009, P = 0.925 > 0.05$). Additionally, no significant interaction was found between Group and Order ($F_{1,75} = 0.208, P = 0.65 > 0.05$). Fig.~\ref{u2_abs}(a) also confirms the conclusions, where we see that participants in \textbf{RL} group have similar absolute accuracy change, compared with participants in \textbf{Random} group. The similar distribution of absolute accuracy delta change (Fig. \ref{u2_abs}(e)) also exhibits no significant difference between the two groups.}

\textcolor{black}{In terms of \textbf{relative} accuracy delta change, no significant effect was found for Order ($F_{1,75} = 1.403, P = 0.24 > 0.05$) or Group ($F_{1,75} = 0.076, P = 0.783 > 0.05$). Additionally, no significant interaction was found between Group and Order ($F_{1,75} = 0.028, P = 0.867 > 0.05$). Similar with \textbf{absolute} results, we find similar \textbf{relative} accuracy change in both groups in the boxplot of Fig.~\ref{u2_rela}(a) and distribution plot of Fig. \ref{u2_rela}(e).
}

\textcolor{black}{The conclusions above are reasonable because we asked participants to always take accuracy as a priority over response time. Therefore, the accuracy of users' choices should be high in most cases, whereas response time varies mainly depending on the time pressure feedback.}

\subsubsection{Attention}

\textcolor{black}{For \textbf{absolute} attention delta change, no significant effect was found for Group ($F_{1,75} = 0.064, P = 0.801 > 0.05$). Interestingly, significant effect was found for Order ($F_{1,75} = 16.685, P < 0.001$), \ie, \textbf{Order 1: Control Session First} (mean $\pm$ SD: $ -0.650 \pm 1.122 $) v.s. \textbf{Order 2: Feedback Session First} (mean $\pm$ SD: $ 0.436 \pm 1.210 $). No significant interaction was found between Group and Order ($F_{1,75} = 0.397, P = 0.530 > 0.05$). Fig.~\ref{u2_abs}(c,g) also support the conclusions, showing that participants have similar \textbf{absolute} attention change and distribution in both groups. 
}

\textcolor{black}{For \textbf{relative} attention change, no significant effect was found for Group ($F_{1,75} = 0.241, P = 0.625 > 0.05$). However, significant effect was found for Order ($F_{1,75} = 18.059, P < 0.001$) as well, \ie \textbf{Order 1: Control Session First} (mean $\pm$ SD: $ -0.125 \pm 0.221 $) v.s. \textbf{Order 2: Feedback Session First} (mean $\pm$ SD: $ 0.170 \pm 0.370 $). No significant interaction was found between Group and Order ($F_{1,75} = 0.218, P = 0.642 > 0.05$). 
Similarly, we also find similar \textbf{relative} attention change and distribution in both groups(Fig. \ref{u2_rela}(c,g)).
}

\textcolor{black}{The significant effect on Order is reasonable as well, since user attention is correlated with the process of the study. For example, user attention may decrease with time. Therefore, if \textbf{Control} session goes first, user attention may drop in the later \textbf{Feedback} session. However, when \textbf{Feedback} session goes first, user attention may drop in the later \textbf{Control} session. Therefore, user attention change is more correlated with the order of the two test sessions.}

\begin{figure}
\centering
\includegraphics[width=1\linewidth]{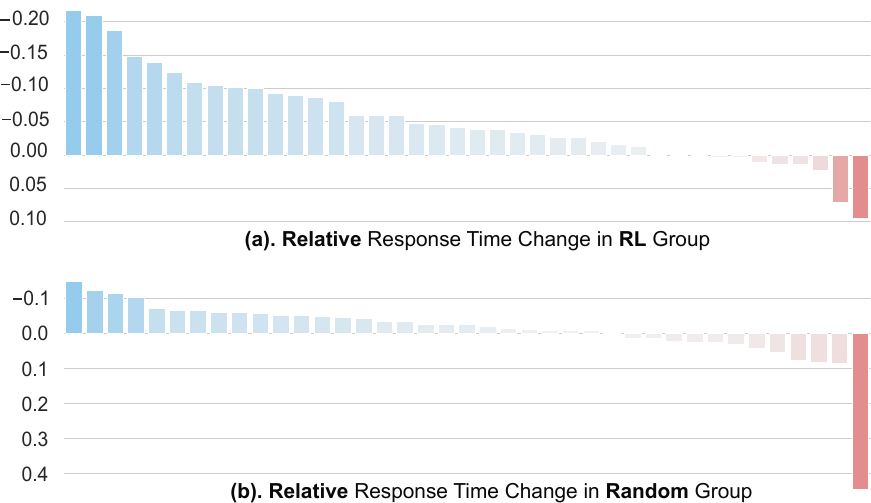}
\caption{\textcolor{black}{\textbf{Relative} response time delta change from \textbf{Control} session to \textbf{Feedback} session for individual participants in two groups (a: \textbf{RL} group, b: \textbf{Random} group). Each bar represents one participant. According to the formula in Table. \ref{tab:metric_calculate}, if the value < 0, it means user response time gets reduced in \textbf{Feedback} session compared with \textbf{Control} session and user performance gets improved. The bar plot is sorted according to the relative response time delta change in two groups.}}
\Description{\textbf{Relative} response time delta change from \textbf{Control} session to \textbf{Feedback} session for individual participants in two groups (a: \textbf{RL} group, b: \textbf{Random} group). Each bar represents one participant. According to the formula in Table. \ref{tab:metric_calculate}, if the value < 0, it means user response time gets reduced in \textbf{Feedback} session compared with \textbf{Control} session and user performance gets improved. The bar plot is sorted according to the relative response time delta change in two groups.}
\label{rela_individual_rl_random}
\end{figure}

\subsubsection{Anxiety}

\textcolor{black}{In terms of \textbf{absolute} anxiety change, no significant effect was found for Order ($F_{1,75} = 0.359, P = 0.551 > 0.05$), Group ($F_{1,75} = 1.301, P = 0.258 > 0.05$), or the interaction between Group and Order ($F_{1,75} = 2.046, P = 0.157 > 0.05$). The similar \textbf{absolute} anxiety change in Fig.~\ref{u2_abs}(d) also corroborates the conclusions above.}

\textcolor{black}{In terms of \textbf{relative} anxiety delta change, no significant effect was found for Order ($F_{1,75} = 1.135, P = 0.29 > 0.05$) or Group ($F_{1,75} = 2.323, P = 0.132 > 0.05$) as well. Interestingly, significant interaction was found between Group and Order ($F_{1,75} = 6.029, P = 0.016 < 0.05$). Fig.~\ref{u2_rela}(d) also depicts the similar \textbf{relative} anxiety change in both groups.
}

\textcolor{black}{Although no significant differences manifest between two groups, we indeed found that the majority of the anxiety level was lower in \textbf{RL} group, compared with that in \textbf{Random} group, as revealed in the distribution plot of both \textbf{absolute} and \textbf{relative} anxiety delta change in Fig.~\ref{u2_abs}(h) and Fig.~\ref{u2_rela}(h).}

\begin{figure}
\centering
\includegraphics[width=1\linewidth]{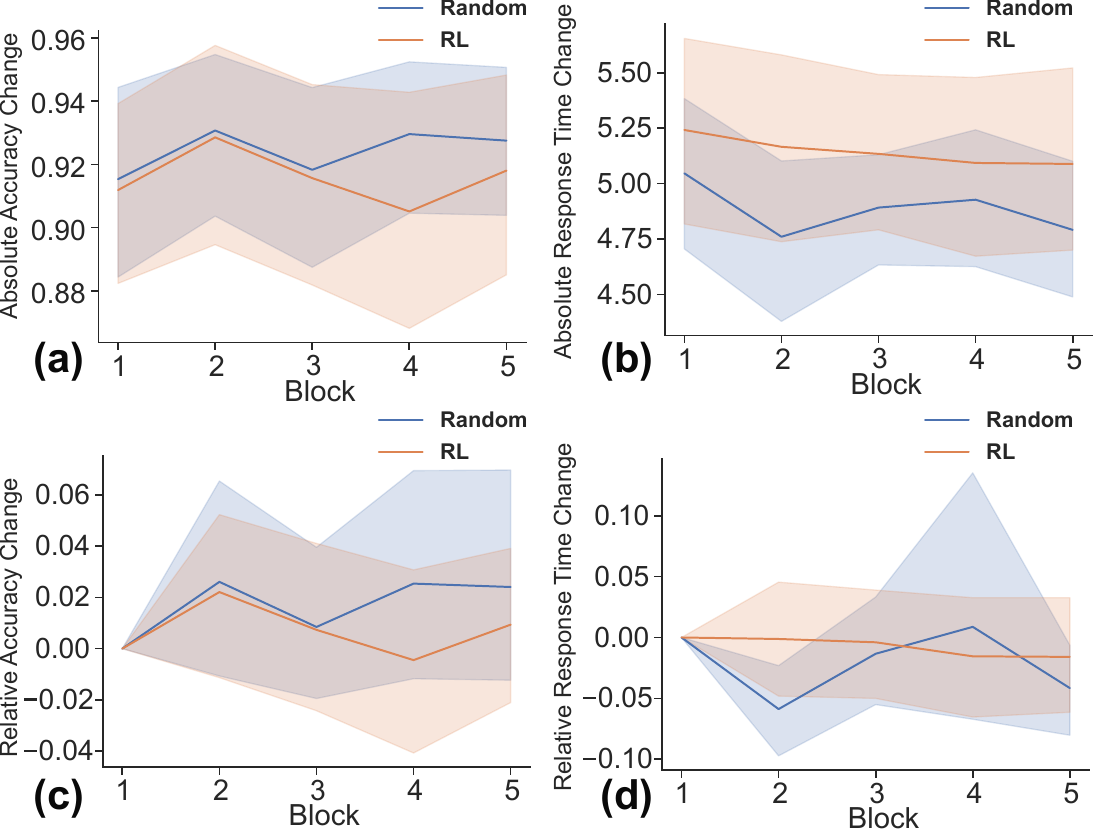}
\caption{\textcolor{black}{\textbf{Absolute} and \textbf{relative} delta change of accuracy (a,c) and response time (b,d) across five blocks in two groups.}}
\Description{\textbf{Absolute} and \textbf{relative} delta change of accuracy (a,c) and response time (b,d) across five blocks in two groups.}
\label{result_trend_compare}
\end{figure}

\subsection{Further Analysis}

We further analyze study results and reveal potential insights to explain how regulation DRL agent could improve user performance.

\subsubsection{Individual User Performance Improvement}

\textcolor{black}{In our previous analysis, we compare user performance improvement between \textbf{RL} and \textbf{Random} group in a two-way between-subjects ANOVA. Now we measure the individual differences for each participant in the two groups so that we could have a more fine-grained understanding of the effect of the regulation DRL agent. Fig.~\ref{resptime_individual} depicts the \textbf{absolute} average response time for each participant in each test session (\textbf{Control}, \textbf{Feedback}) of each group. We observe that RL-based feedback control leads to larger reduction of response time for more participants, compared with random feedback. In terms of \textbf{relative} response time change, Fig.~\ref{rela_individual_rl_random} shows the relative response time delta change from \textbf{Control} session to \textbf{Feedback} session in \textbf{RL} group and \textbf{Random} group, respectively. The results are calculated using the corresponding formula for relative response time delta change in Table. \ref{tab:metric_calculate}. We find that the \textbf{RL} group has larger relative response time reduction from \textbf{Control} session to \textbf{Feedback} session, compared with \textbf{Random} group. \textit{These \textbf{relative} comparison results also corroborate the conclusions in the \textbf{absolute} comparison results}.}

\subsubsection{Trends of User Performance Change}

\textcolor{black}{In order to have a deeper and more fine-grained understanding of user cognition performance change, we break the 100 trials of \textbf{Feedback} session into 5 blocks uniformly in chronological order and investigate the change trend of user cognition performance in both \textbf{RL} and \textbf{Random} group. We calculate the average \textbf{relative} and \textbf{absolute} response time/accuracy in each block. Different from our previous analysis, here we only focus on the \textbf{Feedback} session. Therefore, the \textbf{absolute} response time/accuracy directly mean the average response time/accuracy in each block, \ie, $Block_i\ (1 \leq i \leq 5)$. For \textbf{relative} response time/accuracy, we use average performance of the first block $Block_1$ as the initial performance. In this case, the relative accuracy/response time in other blocks $Block_i\ (2 \leq i \leq 5)$ could be calculated by $(Block_i - Block_1)/Block_1$. }

\textcolor{black}{The \textbf{absolute} and \textbf{relative} accuracy in \textbf{RL} and \textbf{Random} groups are depicted in Fig. \ref{result_trend_compare} (a,c), respectively. No specific trend is found for both \textbf{Random} and \textbf{RL} group. User accuracy may drop or increase with blocks, but this change is not significant enough either, which is also consistent with the ANOVA results. The reason is similar with previous conclusions, \ie, we have asked participants to first take accuracy as the priority. Therefore, user response time may vary but accuracy will not. }

\textcolor{black}{In terms of response time, both \textbf{absolute} and \textbf{relative} response time in two groups are depicted in Fig. \ref{result_trend_compare} (b,d), respectively. We could find that \textbf{RL} group has obvious and consistent response time reduction, which is not found in \textbf{Random} group. Instead, response time in \textbf{Random} group may decrease at first but increase again in the later blocks, which does not have stable response time reduction. This phenomenon corroborates our previous conclusions and demonstrates the superiority of RL-based time pressure feedback.}

\subsubsection{Visualizing the Time Pressure Feedback Trajectory}
\label{sec: trajectory visual}

\begin{figure}
\centering
\includegraphics[width=0.8\linewidth]{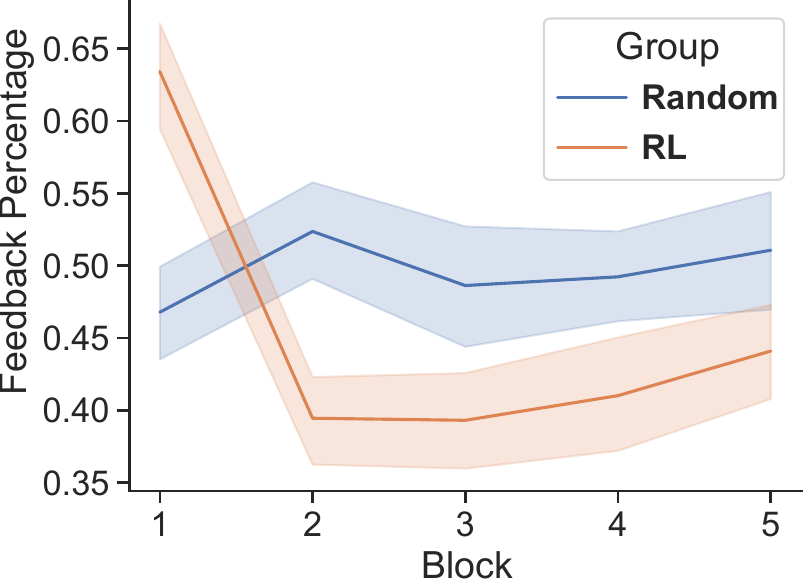}
\caption{\textcolor{black}{Time pressure feedback percentage in five blocks in \textbf{RL} group and \textbf{Random} group.}}
\Description{Time pressure feedback percentage in five blocks in \textbf{RL} group and \textbf{Random} group.}
\label{feedback_trend}
\end{figure}

\textcolor{black}{In addition to analyzing the average response time/accuracy during each block, we also measure the proportion of trials under time pressure feedback with respect to all trials of each block (named as feedback percentage in each block). By doing so, we could compare and analyze the trajectory of time pressure feedback along with different blocks in two groups, as visualized in Fig. \ref{feedback_trend}. }

\textcolor{black}{For \textbf{Random} group, we could find that the feedback percentage is always around 50\% in each block. This is consistent with our definition of random feedback, where there is 50\% chance that the time pressure feedback may happen during each trial. This is similar across all five blocks in \textbf{Random} group.}

\textcolor{black}{For \textbf{RL} group, we could find that the feedback trajectory, or the feedback percentage change trend across five blocks, is totally different from the \textbf{Random} group. \textit{The feedback percentage is larger than 50\% at first in $Block_1$ and then decreases immediately in the later blocks}. However, the feedback percentage also increases slowly from $Block_2$ to $Block_5$. This is probably due to the adaptive ability of the regulation DRL agent. Specifically, when time pressure leads to worse user performance, the regulation DRL agent will decrease the proportion of time pressure feedback. However, if the time pressure does not worsen user performance, it will be activated again to raise users' arousal level and regulate user cognition performance.}

\textcolor{black}{In summary, the trajectory of time pressure feedback also reveals the adaptive and far-sighted ability of the regulation DRL agent to improve the overall user cognition performance. }

\subsubsection{User Subjective Experience}

We also examined the users' subjective experience through an informal interview after the user study. Specifically, we asked participants about their feelings during the study when time pressure feedback happened or disappeared. Overall, most users admitted that the time pressure feedback indeed introduced pressure and conveyed a sense of urgency to them during the task. \textcolor{black}{One participant said that time pressure feedback motivated him to answer questions faster but it also increased his anxiety. Interestingly, although participants knew that there was no penalty associated with the progress bar, they would still try their best to answer the question before the progress bar achieved $100\%$. This made them answer questions faster but also feel more stressed. Both the advantages and drawbacks of time pressure feedback from participants' subjective experience demonstrate the necessity to provide flexible time pressure feedback that could adaptively control the existence of time pressure feedback according to user real-time performance. }

\textcolor{black}{Additionally, although participants admitted the advantage of time pressure to make them finish math questions faster, one participant in \textbf{Random} group said that the time pressure prevented him from focusing on the math task. Another participant also stated that the random time pressure was not constant and would just introduce too much distraction than anything else to him. }

\textcolor{black}{However, one participant in \textbf{RL} group said that although the time pressure made her lose focus and feel anxious in the beginning, after the progress bar popped up for a few more times, she got more accustomed to it and became less anxious in the later trials. This may reveal the adaptability of the regulation DRL agent to control the time pressure feedback adaptively to implicitly mitigate the distractive artifacts compared with the random feedback strategy.}

\textcolor{black}{What's more, one participant in our previous pilot study also said that she felt more focused in the beginning so time pressure feedback did not distract her too much. But as the study went on, she felt tired and time pressure feedback actually introduced more anxiety for her. This situation echoes the conclusions from the time pressure feedback trajectory visualization in Sec. \ref{sec: trajectory visual}, where we find that time pressure percentage drops after $Block_1$ in \textbf{RL} group. We could infer that the regulation DRL agent has probably learned that time pressure feedback may introduce more anxiety to users compared with the beginning of the study in the simulation environment. Therefore, in the real user study, less time pressure feedback was provided to participants after $Block_1$. }

In summary, \textit{the subjective experiences of participants also match our conclusions from the study and demonstrate the necessity and effectiveness of adaptive time pressure feedback controlled by the regulation DRL agent}.

\section{Potential Application Scenarios}

\textcolor{black}{\textcolor{black}{In this paper}, we use math arithmetic task as the cognition task and visual progress bar as the visual stimuli to investigate and explore our AI-based human cognition augmentation framework. Admittedly, these represent relatively simple and controlled settings, which may not replicate the real-world scenarios. However, we believe this marks an initial step towards the vision of harnessing AI to augment human cognition. Our approach may inspire practical application scenarios below.
}

\subsection{Improving Working Efficiency}

\textcolor{black}{Existing research has explored various technologies to improve working efficiency. For instance, by controlling the time rate displayed by a virtual clock, user working productivity could be further improved \cite{ban2015improving}. In addition, the working efficiency of people with disabilities could be improved by applying the universal design concept for working environment enhancement \cite{pruettikomon2018study}. Moreover, the recent advances of ubiquitous and mobile computing could now support working performance prediction \cite{mirjafari2021predicting} and working efficiency enhancement \cite{10.1145/2037373.2037501}. Furthermore, AI could even improve working efficiency via adaptive guidance of farming works considering physical behaviors in a machine learning model\cite{10.1145/3055635.3056616}. Such work demonstrates the feasibility of harnessing AI-mediated intervention to improve working efficiency. For example, although the popular \textcolor{black}{Pomodoro} Timer software \cite{shammas2019simple} may increase user working productivity, it may also introduce additional burden and anxiety to users because users could always feel the urgent sense of passing time. Adaptive control of \textcolor{black}{the Pomodoro} timer may solve this problem to allow the existence of the timer when a user is absent-minded and hide it when a user is anxious. 
Similar adaptive feedback patterns could be utilized to augment user working efficiency, which is not limited by time pressure feedback or a \textcolor{black}{Pomodoro} timer. }

\subsection{Improving Learning Outcome}

\textcolor{black}{Student learning outcome could be improved by computer-based technologies \cite{10.1145/3159450.3159500}, interactive learning styles \cite{10.5555/1231091.1231096}, and multiple teaching modalities \cite{10.1145/3328778.3366880}.  In addition, virtual communities and mobile platforms could foster student engagement to achieve better learning outcomes \cite{10.1145/3084381.3084434}. Furthermore, ubiquitous sensing in the classroom \cite{10.1145/3411764.3445711} such as EduSense\cite{10.1145/3351229} and n-Gage\cite{10.1145/3411813} could provide more insightful information to instructors and therefore improve learning outcome. Similar with our framework to augment user cognition, existing work has also demonstrated the feasibility of leveraging reinforcement learning models to adaptively schedule educational activities and improve student learning performance \cite{bassen2020reinforcement}. Based on these existing exploration, we believe that it is feasible and promising to utilize a similar AI-mediated framework to augment learning outcome via adaptive intervention control.  }

\subsection{Adaptive Mental State Regulation}

\textcolor{black}{Existing biofeedback systems usually integrate sensing and feedback modalities to regulate user mental states. For instance, MoodWings \cite{maclean2013moodwings} could allow users to self-regulate their stress status via mirroring users' real-time stress state through actuated wing motion. In addition, EmotionCheck leverages bodily signals and false heart-beat feedback to regulate user emotion \cite{costa2016emotioncheck}. By contrast, BoostMeUp\cite{costa2019boostmeup} generates personalized haptic feedback in a different heart rate to regulate user emotion and improve cognitive performance. False heart-rate feedback is also verified to be able to create an interoceptive illusion of effort on users\cite{iodice2019interoceptive}. Moreover, online neurofeedback could also regulate user arousal and therefore improve human performance\cite{faller2019regulation}. Finally, biofeedback systems like AttentivU \cite{kosmyna2018attentivu} could integrate EEG to provide adaptive intervention to regulate user emotion. Such real-time intervention systems either simply present users' current mental states to facilitate self-regulation, or provide adaptive but simple first-order heuristics feedback to regulate users' mental states. These myopic strategies may not always improve user engagement but may act as a distraction. Based on our framework, AI-mediated intelligent strategies could further improve the adaptation of biofeedback systems to regulate user mental states in a far-sighted manner. }

\section{Discussion, Limitations, and Future Work}

\subsection{Opportunities and Challenges}

In this work, we leverage the pre-trained regulation DRL agent in simulation environment to regulate cognition performance of real users without tailoring the model to specific users. This framework design solves three main problems of straightforward online DRL training in real user studies. First, it is hard to obtain adequate data for DRL agent training in human-involved model training process. Second, fine-tuning hyperparameters of the DRL agent may require iterative user studies. Third, the initial exploration of DRL agent in training process may be quite random and user performance may get very bad if this initial training process directly happens in real users. 
\textcolor{black}{Existing work has also shown the effectiveness of leveraging reinforcement learning to regulate human behaviors in various scenarios such as education \cite{bassen2020reinforcement}, human-machine interaction for sound design \cite{scurto2021designing}, health monitoring and intervention \cite{liao2020personalized}, etc. Our work shares similar spirit especially in the human-in-the-loop reinforcement learning. However, our work harnesses a dual-DRL framework to deal with the insufficient training data problem in the human-in-the-loop training process, while others either use a simplified algorithm \cite{scurto2021designing,liao2020personalized} to reduce the requirement of mass training data or leverage existing simulation environment \cite{bassen2020reinforcement}. The future work could focus on the exploration and comparison of these different methodologies to tackle the problem of data insufficiency.}

\textcolor{black}{
It is important to note that RL-based models may fail in certain circumstances and there are still many challenges\cite{10.1145/3527448}. For instance, the transfer from simulation to real environments is usually not perfect \cite{10.1145/3447548.3467181}, which may deteriorate the performance of RL models. In addition, a general model may not represent all users. For example, in our user study results, there are still few participants whose response time is not improved by the regulation DRL agent. In this case, a personalized model may potentially solve this problem. Therefore , we could fine-tune the pre-trained regulation DRL agent for each real user to further improve the cognitive performance. 
}

\subsection{Limitations of the User Study}

\textcolor{black}{Although we have demonstrated the feasibility of our framework, a few constraints need to be taken into account before studying its broader implications.}

\subsubsection{Study Population}

\textcolor{black}{Our user study recruited 80 participants to run a between-subjects evaluation. However, a larger-scale user study may be helpful to further verify the effectiveness of our framework. \textcolor{black}{The reason is that the participants we recruited are mostly young people (age: around 20 years) with similar education background (undergraduates and graduates). A larger and more diverse group of participants may reveal new insights and findings.} 
In addition, we may also explore the effectiveness of the regulation DRL agent in the long-term training of user cognition performance.
}

\subsubsection{Baseline Group Selection}

\textcolor{black}{In this work, we use \textbf{Random} group to compare with \textbf{RL} group. The reason 
has been clarified in Section \ref{sec: experiment design}. However, it is possible that there might be more advanced feedback control strategies,  \textcolor{black}{e.g., Deep Tamer\cite{warnell2018deep} and Deep COACH\cite{arumugam2019deep}. A comparison between these models and our framework may reveal new insights}. We expect the open-source of our software/dataset can inspire the development of such models.}

\subsubsection{Study Results and Findings}

\textcolor{black}{Our user study uses a two-way between-subjects ANOVA to demonstrate the significant differences between \textbf{Random} and \textbf{RL} group in response time change from \textbf{Control} session to \textbf{Feedback} session. However, we did not find statistically significant differences in user emotion change including attention and anxiety change between the two groups, although differences exist in the attention/anxiety score distribution. One reason may be that participants just give subjective scores about their feelings, which may vary significantly across different participants. Therefore, it is hard to normalize different participants' feelings.  
A more standard emotion evaluation questionnaire may have the potential to solve this problem such as the State-Trait Anxiety Inventory (STAI) questionnaire \cite{spielberger1971state}. 
However, such evaluation consists of a large number of questions, which may stress the participants and affect the results. A better way to explore user emotion change could be our future work.}

\subsection{Generalization to Different Cognition Tasks and Feedback}
As discussed in Sec. \ref{sec:intro}, the math arithmetic task and time pressure feedback are used as example modalities to demonstrate the feasibility of our vision of augmenting human cognition using AI models.
In general, for other cognition tasks, we could similarly train a corresponding simulation DRL agent to serve as the simulation environment. This is feasible since existing work has demonstrated the power of neural network models in solving many cognition tasks\cite{yang2019task}. Given the simulation DRL agent, we could similarly train a regulation DRL agent to interact with the simulation environment. The trained regulation DRL agent can then be applied to regulate real users. 
Although we may need to collect new datasets for new cognition tasks to train the simulation environment, it is also possible to directly use the pre-trained regulation DRL agent in math arithmetic task to regulate users in other cognition tasks. The reason is that the regulation DRL agent only takes user cognition results (response time) as observation space. It will not need cognition task information as input. The regulation DRL agent may capture potential correlation between time pressure feedback control and user response time regardless of the task. 
We leave such generalization for future work.

In addition, our framework could also be extended to different feedback modalities. Our current work uses visual time pressure feedback to regulate user performance, which is also used in the observation space of simulation DRL agent and action space of the regulation DRL agent. For other kinds of visual feedback, we could similarly use the corresponding visual information in the format of image/video.
The format can be updated for other feedback modalities. For example, for auditory feedback, we could use digital audio signals instead of image/video in the observation space of simulation DRL agent and the action space of regulation DRL agent.

\section{Conclusion}

In this paper, we present our exploration in intelligent feedback to augment human cognition performance by developing adaptive time pressure visual feedback in a math arithmetic task. The control strategy is based on a regulation DRL agent, which is pre-trained by interacting with another simulation DRL agent. In a user study, we demonstrate the feasibility and effectiveness of this framework design and provide explanation about how the regulation DRL agent could improve user cognition performance. \textcolor{black}{We also discuss potential application scenarios, limitations, and future work to increase the scalability of this work and make larger impact into different research areas.} We believe that our work could serve as important groundwork for future exploration of human ability augmentation with intelligent feedback.

\bibliographystyle{ACM-Reference-Format}
\bibliography{main}


\begin{thebibliography}{74}


\ifx \showCODEN    \undefined \def \showCODEN     #1{\unskip}     \fi
\ifx \showDOI      \undefined \def \showDOI       #1{#1}\fi
\ifx \showISBNx    \undefined \def \showISBNx     #1{\unskip}     \fi
\ifx \showISBNxiii \undefined \def \showISBNxiii  #1{\unskip}     \fi
\ifx \showISSN     \undefined \def \showISSN      #1{\unskip}     \fi
\ifx \showLCCN     \undefined \def \showLCCN      #1{\unskip}     \fi
\ifx \shownote     \undefined \def \shownote      #1{#1}          \fi
\ifx \showarticletitle \undefined \def \showarticletitle #1{#1}   \fi
\ifx \showURL      \undefined \def \showURL       {\relax}        \fi
\providecommand\bibfield[2]{#2}
\providecommand\bibinfo[2]{#2}
\providecommand\natexlab[1]{#1}
\providecommand\showeprint[2][]{arXiv:#2}

\bibitem[Ahuja et~al\mbox{.}(2019)]%
        {10.1145/3351229}
\bibfield{author}{\bibinfo{person}{Karan Ahuja}, \bibinfo{person}{Dohyun Kim},
  \bibinfo{person}{Franceska Xhakaj}, \bibinfo{person}{Virag Varga},
  \bibinfo{person}{Anne Xie}, \bibinfo{person}{Stanley Zhang},
  \bibinfo{person}{Jay~Eric Townsend}, \bibinfo{person}{Chris Harrison},
  \bibinfo{person}{Amy Ogan}, {and} \bibinfo{person}{Yuvraj Agarwal}.}
  \bibinfo{year}{2019}\natexlab{}.
\newblock \showarticletitle{EduSense: Practical Classroom Sensing at Scale}.
\newblock \bibinfo{journal}{\emph{Proc. ACM Interact. Mob. Wearable Ubiquitous
  Technol.}} \bibinfo{volume}{3}, \bibinfo{number}{3}, Article
  \bibinfo{articleno}{71} (\bibinfo{date}{sep} \bibinfo{year}{2019}),
  \bibinfo{numpages}{26}~pages.
\newblock
\urldef\tempurl%
\url{https://doi.org/10.1145/3351229}
\showDOI{\tempurl}


\bibitem[Ahuja et~al\mbox{.}(2021)]%
        {10.1145/3411764.3445711}
\bibfield{author}{\bibinfo{person}{Karan Ahuja}, \bibinfo{person}{Deval Shah},
  \bibinfo{person}{Sujeath Pareddy}, \bibinfo{person}{Franceska Xhakaj},
  \bibinfo{person}{Amy Ogan}, \bibinfo{person}{Yuvraj Agarwal}, {and}
  \bibinfo{person}{Chris Harrison}.} \bibinfo{year}{2021}\natexlab{}.
\newblock \showarticletitle{Classroom Digital Twins with Instrumentation-Free
  Gaze Tracking}. In \bibinfo{booktitle}{\emph{Proceedings of the 2021 CHI
  Conference on Human Factors in Computing Systems}} (Yokohama, Japan)
  \emph{(\bibinfo{series}{CHI '21})}. \bibinfo{publisher}{Association for
  Computing Machinery}, \bibinfo{address}{New York, NY, USA}, Article
  \bibinfo{articleno}{484}, \bibinfo{numpages}{9}~pages.
\newblock
\showISBNx{9781450380966}
\urldef\tempurl%
\url{https://doi.org/10.1145/3411764.3445711}
\showDOI{\tempurl}


\bibitem[Arumugam et~al\mbox{.}(2019)]%
        {arumugam2019deep}
\bibfield{author}{\bibinfo{person}{Dilip Arumugam}, \bibinfo{person}{Jun~Ki
  Lee}, \bibinfo{person}{Sophie Saskin}, {and} \bibinfo{person}{Michael~L
  Littman}.} \bibinfo{year}{2019}\natexlab{}.
\newblock \showarticletitle{Deep reinforcement learning from policy-dependent
  human feedback}.
\newblock \bibinfo{journal}{\emph{arXiv preprint arXiv:1902.04257}}
  (\bibinfo{year}{2019}).
\newblock


\bibitem[Baham\'{o}n and Rorrer(2020)]%
        {10.1145/3328778.3366880}
\bibfield{author}{\bibinfo{person}{Julio~C\'{e}sar Baham\'{o}n} {and}
  \bibinfo{person}{Audrey Rorrer}.} \bibinfo{year}{2020}\natexlab{}.
\newblock \showarticletitle{Improving Student Learning Outcomes in Online
  Courses: An Investigation Into the Effects of Multiple Teaching Modalities}.
  In \bibinfo{booktitle}{\emph{Proceedings of the 51st ACM Technical Symposium
  on Computer Science Education}} (Portland, OR, USA)
  \emph{(\bibinfo{series}{SIGCSE '20})}. \bibinfo{publisher}{Association for
  Computing Machinery}, \bibinfo{address}{New York, NY, USA},
  \bibinfo{pages}{1179–1185}.
\newblock
\showISBNx{9781450367936}
\urldef\tempurl%
\url{https://doi.org/10.1145/3328778.3366880}
\showDOI{\tempurl}


\bibitem[Ban et~al\mbox{.}(2015)]%
        {ban2015improving}
\bibfield{author}{\bibinfo{person}{Yuki Ban}, \bibinfo{person}{Sho Sakurai},
  \bibinfo{person}{Takuji Narumi}, \bibinfo{person}{Tomohiro Tanikawa}, {and}
  \bibinfo{person}{Michitaka Hirose}.} \bibinfo{year}{2015}\natexlab{}.
\newblock \showarticletitle{Improving work productivity by controlling the time
  rate displayed by the virtual clock}. In
  \bibinfo{booktitle}{\emph{Proceedings of the 6th Augmented Human
  International Conference}}. \bibinfo{pages}{25--32}.
\newblock


\bibitem[Bassen et~al\mbox{.}(2020)]%
        {bassen2020reinforcement}
\bibfield{author}{\bibinfo{person}{Jonathan Bassen}, \bibinfo{person}{Bharathan
  Balaji}, \bibinfo{person}{Michael Schaarschmidt}, \bibinfo{person}{Candace
  Thille}, \bibinfo{person}{Jay Painter}, \bibinfo{person}{Dawn Zimmaro},
  \bibinfo{person}{Alex Games}, \bibinfo{person}{Ethan Fast}, {and}
  \bibinfo{person}{John~C Mitchell}.} \bibinfo{year}{2020}\natexlab{}.
\newblock \showarticletitle{Reinforcement learning for the adaptive scheduling
  of educational activities}. In \bibinfo{booktitle}{\emph{Proceedings of the
  2020 CHI Conference on Human Factors in Computing Systems}}.
  \bibinfo{pages}{1--12}.
\newblock


\bibitem[Battleday et~al\mbox{.}(2017)]%
        {battleday2017modeling}
\bibfield{author}{\bibinfo{person}{Ruairidh~M Battleday},
  \bibinfo{person}{Joshua~C Peterson}, {and} \bibinfo{person}{Thomas~L
  Griffiths}.} \bibinfo{year}{2017}\natexlab{}.
\newblock \showarticletitle{Modeling human categorization of natural images
  using deep feature representations}.
\newblock \bibinfo{journal}{\emph{arXiv preprint arXiv:1711.04855}}
  (\bibinfo{year}{2017}).
\newblock


\bibitem[Battleday et~al\mbox{.}(2020)]%
        {battleday2020capturing}
\bibfield{author}{\bibinfo{person}{Ruairidh~M Battleday},
  \bibinfo{person}{Joshua~C Peterson}, {and} \bibinfo{person}{Thomas~L
  Griffiths}.} \bibinfo{year}{2020}\natexlab{}.
\newblock \showarticletitle{Capturing human categorization of natural images by
  combining deep networks and cognitive models}.
\newblock \bibinfo{journal}{\emph{Nature communications}} \bibinfo{volume}{11},
  \bibinfo{number}{1} (\bibinfo{year}{2020}), \bibinfo{pages}{1--14}.
\newblock


\bibitem[Battleday et~al\mbox{.}(2021)]%
        {battleday2021convolutional}
\bibfield{author}{\bibinfo{person}{Ruairidh~M Battleday},
  \bibinfo{person}{Joshua~C Peterson}, {and} \bibinfo{person}{Thomas~L
  Griffiths}.} \bibinfo{year}{2021}\natexlab{}.
\newblock \showarticletitle{From convolutional neural networks to models of
  higher-level cognition (and back again)}.
\newblock \bibinfo{journal}{\emph{Annals of the New York Academy of Sciences}}
  \bibinfo{volume}{1505}, \bibinfo{number}{1} (\bibinfo{year}{2021}),
  \bibinfo{pages}{55--78}.
\newblock


\bibitem[Bourgin et~al\mbox{.}(2019)]%
        {bourgin2019cognitive}
\bibfield{author}{\bibinfo{person}{David~D Bourgin}, \bibinfo{person}{Joshua~C
  Peterson}, \bibinfo{person}{Daniel Reichman}, \bibinfo{person}{Stuart~J
  Russell}, {and} \bibinfo{person}{Thomas~L Griffiths}.}
  \bibinfo{year}{2019}\natexlab{}.
\newblock \showarticletitle{Cognitive model priors for predicting human
  decisions}. In \bibinfo{booktitle}{\emph{International conference on machine
  learning}}. PMLR, \bibinfo{pages}{5133--5141}.
\newblock


\bibitem[Brice and Smith(2002)]%
        {brice2002effects}
\bibfield{author}{\bibinfo{person}{Carolyn~F Brice} {and}
  \bibinfo{person}{Andrew~P Smith}.} \bibinfo{year}{2002}\natexlab{}.
\newblock \showarticletitle{Effects of caffeine on mood and performance: a
  study of realistic consumption}.
\newblock \bibinfo{journal}{\emph{Psychopharmacology}} \bibinfo{volume}{164},
  \bibinfo{number}{2} (\bibinfo{year}{2002}), \bibinfo{pages}{188--192}.
\newblock


\bibitem[Brockman et~al\mbox{.}(2016)]%
        {brockman2016openai}
\bibfield{author}{\bibinfo{person}{Greg Brockman}, \bibinfo{person}{Vicki
  Cheung}, \bibinfo{person}{Ludwig Pettersson}, \bibinfo{person}{Jonas
  Schneider}, \bibinfo{person}{John Schulman}, \bibinfo{person}{Jie Tang},
  {and} \bibinfo{person}{Wojciech Zaremba}.} \bibinfo{year}{2016}\natexlab{}.
\newblock \showarticletitle{Openai gym}.
\newblock \bibinfo{journal}{\emph{arXiv preprint arXiv:1606.01540}}
  (\bibinfo{year}{2016}).
\newblock


\bibitem[Butt and Sultan(2011)]%
        {butt2011coffee}
\bibfield{author}{\bibinfo{person}{Masood~Sadiq Butt} {and}
  \bibinfo{person}{M~Tauseef Sultan}.} \bibinfo{year}{2011}\natexlab{}.
\newblock \showarticletitle{Coffee and its consumption: benefits and risks}.
\newblock \bibinfo{journal}{\emph{Critical reviews in food science and
  nutrition}} \bibinfo{volume}{51}, \bibinfo{number}{4} (\bibinfo{year}{2011}),
  \bibinfo{pages}{363--373}.
\newblock


\bibitem[Chen et~al\mbox{.}(2017)]%
        {chen2017cognitive}
\bibfield{author}{\bibinfo{person}{Xiuli Chen},
  \bibinfo{person}{Sandra~Dorothee Starke}, \bibinfo{person}{Chris Baber},
  {and} \bibinfo{person}{Andrew Howes}.} \bibinfo{year}{2017}\natexlab{}.
\newblock \showarticletitle{A cognitive model of how people make decisions
  through interaction with visual displays}. In
  \bibinfo{booktitle}{\emph{Proceedings of the 2017 CHI conference on human
  factors in computing systems}}. \bibinfo{pages}{1205--1216}.
\newblock


\bibitem[Cheng(2017)]%
        {cheng2017evaluation}
\bibfield{author}{\bibinfo{person}{Shyh-Yueh Cheng}.}
  \bibinfo{year}{2017}\natexlab{}.
\newblock \showarticletitle{Evaluation of effect on cognition response to time
  pressure by using EEG}. In \bibinfo{booktitle}{\emph{International conference
  on applied human factors and ergonomics}}. Springer, \bibinfo{pages}{45--52}.
\newblock


\bibitem[Childs(2014)]%
        {childs2014influence}
\bibfield{author}{\bibinfo{person}{Emma Childs}.}
  \bibinfo{year}{2014}\natexlab{}.
\newblock \showarticletitle{Influence of energy drink ingredients on mood and
  cognitive performance}.
\newblock \bibinfo{journal}{\emph{Nutrition reviews}} \bibinfo{volume}{72},
  \bibinfo{number}{suppl\_1} (\bibinfo{year}{2014}), \bibinfo{pages}{48--59}.
\newblock


\bibitem[Chollet et~al\mbox{.}(2015)]%
        {chollet2015keras}
\bibfield{author}{\bibinfo{person}{Fran\c{c}ois Chollet} {et~al\mbox{.}}}
  \bibinfo{year}{2015}\natexlab{}.
\newblock \bibinfo{title}{Keras}.
\newblock \bibinfo{howpublished}{\url{https://keras.io}}.
\newblock


\bibitem[Corey(2003)]%
        {corey2003memory}
\bibfield{author}{\bibinfo{person}{Vicka~R Corey}.}
  \bibinfo{year}{2003}\natexlab{}.
\newblock \showarticletitle{The memory glasses: subliminal vs. overt memory
  support with imperfect information}. In \bibinfo{booktitle}{\emph{Proceedings
  of the Seventh IEEE International Symposium on Wearable Computers
  (ISWC’03)}}, Vol.~\bibinfo{volume}{1530}. Citeseer,
  \bibinfo{pages}{17--00}.
\newblock


\bibitem[Costa et~al\mbox{.}(2016)]%
        {costa2016emotioncheck}
\bibfield{author}{\bibinfo{person}{Jean Costa}, \bibinfo{person}{Alexander~T
  Adams}, \bibinfo{person}{Malte~F Jung}, \bibinfo{person}{Fran{\c{c}}ois
  Guimbreti{\`e}re}, {and} \bibinfo{person}{Tanzeem Choudhury}.}
  \bibinfo{year}{2016}\natexlab{}.
\newblock \showarticletitle{EmotionCheck: leveraging bodily signals and false
  feedback to regulate our emotions}. In \bibinfo{booktitle}{\emph{Proceedings
  of the 2016 ACM international joint conference on pervasive and ubiquitous
  computing}}. \bibinfo{pages}{758--769}.
\newblock


\bibitem[Costa et~al\mbox{.}(2019)]%
        {costa2019boostmeup}
\bibfield{author}{\bibinfo{person}{Jean Costa}, \bibinfo{person}{Fran{\c{c}}ois
  Guimbreti{\`e}re}, \bibinfo{person}{Malte~F Jung}, {and}
  \bibinfo{person}{Tanzeem Choudhury}.} \bibinfo{year}{2019}\natexlab{}.
\newblock \showarticletitle{Boostmeup: Improving cognitive performance in the
  moment by unobtrusively regulating emotions with a smartwatch}.
\newblock \bibinfo{journal}{\emph{Proceedings of the ACM on Interactive,
  Mobile, Wearable and Ubiquitous Technologies}} \bibinfo{volume}{3},
  \bibinfo{number}{2} (\bibinfo{year}{2019}), \bibinfo{pages}{1--23}.
\newblock


\bibitem[Cremer and Kasparov(2021)]%
        {hbr}
\bibfield{author}{\bibinfo{person}{David~De Cremer} {and}
  \bibinfo{person}{Garry Kasparov}.} \bibinfo{year}{2021}\natexlab{}.
\newblock \bibinfo{title}{AI Should Augment Human Intelligence, Not Replace
  It}.
\newblock
\newblock
\urldef\tempurl%
\url{https://hbr.org/2021/03/ai-should-augment-human-intelligence-not-replace-it}
\showURL{%
Retrieved january 20, 2023 from \tempurl}


\bibitem[Daitch et~al\mbox{.}(2016)]%
        {daitch2016mapping}
\bibfield{author}{\bibinfo{person}{Amy~L Daitch}, \bibinfo{person}{Brett~L
  Foster}, \bibinfo{person}{Jessica Schrouff}, \bibinfo{person}{Vinitha
  Rangarajan}, \bibinfo{person}{It{\i}r Ka{\c{s}}ik{\c{c}}i},
  \bibinfo{person}{Sandra Gattas}, {and} \bibinfo{person}{Josef Parvizi}.}
  \bibinfo{year}{2016}\natexlab{}.
\newblock \showarticletitle{Mapping human temporal and parietal neuronal
  population activity and functional coupling during mathematical cognition}.
\newblock \bibinfo{journal}{\emph{Proceedings of the National Academy of
  Sciences}} \bibinfo{volume}{113}, \bibinfo{number}{46}
  (\bibinfo{year}{2016}), \bibinfo{pages}{E7277--E7286}.
\newblock


\bibitem[Do et~al\mbox{.}(2021)]%
        {do2021simulation}
\bibfield{author}{\bibinfo{person}{Seungwon Do}, \bibinfo{person}{Minsuk
  Chang}, {and} \bibinfo{person}{Byungjoo Lee}.}
  \bibinfo{year}{2021}\natexlab{}.
\newblock \showarticletitle{A simulation model of intermittently controlled
  point-and-click behaviour}. In \bibinfo{booktitle}{\emph{Proceedings of the
  2021 CHI Conference on Human Factors in Computing Systems}}.
  \bibinfo{pages}{1--17}.
\newblock


\bibitem[Edland and Svenson(1993)]%
        {edland1993judgment}
\bibfield{author}{\bibinfo{person}{Anne Edland} {and} \bibinfo{person}{Ola
  Svenson}.} \bibinfo{year}{1993}\natexlab{}.
\newblock \showarticletitle{Judgment and decision making under time pressure}.
\newblock In \bibinfo{booktitle}{\emph{Time pressure and stress in human
  judgment and decision making}}. \bibinfo{publisher}{Springer},
  \bibinfo{pages}{27--40}.
\newblock


\bibitem[Erev et~al\mbox{.}(2017)]%
        {erev2017anomalies}
\bibfield{author}{\bibinfo{person}{Ido Erev}, \bibinfo{person}{Eyal Ert},
  \bibinfo{person}{Ori Plonsky}, \bibinfo{person}{Doron Cohen}, {and}
  \bibinfo{person}{Oded Cohen}.} \bibinfo{year}{2017}\natexlab{}.
\newblock \showarticletitle{From anomalies to forecasts: Toward a descriptive
  model of decisions under risk, under ambiguity, and from experience.}
\newblock \bibinfo{journal}{\emph{Psychological review}} \bibinfo{volume}{124},
  \bibinfo{number}{4} (\bibinfo{year}{2017}), \bibinfo{pages}{369}.
\newblock


\bibitem[Faller et~al\mbox{.}(2019)]%
        {faller2019regulation}
\bibfield{author}{\bibinfo{person}{Josef Faller}, \bibinfo{person}{Jennifer
  Cummings}, \bibinfo{person}{Sameer Saproo}, {and} \bibinfo{person}{Paul
  Sajda}.} \bibinfo{year}{2019}\natexlab{}.
\newblock \showarticletitle{Regulation of arousal via online neurofeedback
  improves human performance in a demanding sensory-motor task}.
\newblock \bibinfo{journal}{\emph{Proceedings of the National Academy of
  Sciences}} \bibinfo{volume}{116}, \bibinfo{number}{13}
  (\bibinfo{year}{2019}), \bibinfo{pages}{6482--6490}.
\newblock


\bibitem[Fudenberg et~al\mbox{.}(2020)]%
        {fudenberg2020testing}
\bibfield{author}{\bibinfo{person}{Drew Fudenberg}, \bibinfo{person}{Whitney
  Newey}, \bibinfo{person}{Philipp Strack}, {and} \bibinfo{person}{Tomasz
  Strzalecki}.} \bibinfo{year}{2020}\natexlab{}.
\newblock \showarticletitle{Testing the drift-diffusion model}.
\newblock \bibinfo{journal}{\emph{Proceedings of the National Academy of
  Sciences}} \bibinfo{volume}{117}, \bibinfo{number}{52}
  (\bibinfo{year}{2020}), \bibinfo{pages}{33141--33148}.
\newblock


\bibitem[Gao et~al\mbox{.}(2020)]%
        {10.1145/3411813}
\bibfield{author}{\bibinfo{person}{Nan Gao}, \bibinfo{person}{Wei Shao},
  \bibinfo{person}{Mohammad~Saiedur Rahaman}, {and} \bibinfo{person}{Flora~D.
  Salim}.} \bibinfo{year}{2020}\natexlab{}.
\newblock \showarticletitle{N-Gage: Predicting in-Class Emotional, Behavioural
  and Cognitive Engagement in the Wild}.
\newblock \bibinfo{journal}{\emph{Proc. ACM Interact. Mob. Wearable Ubiquitous
  Technol.}} \bibinfo{volume}{4}, \bibinfo{number}{3}, Article
  \bibinfo{articleno}{79} (\bibinfo{date}{sep} \bibinfo{year}{2020}),
  \bibinfo{numpages}{26}~pages.
\newblock
\urldef\tempurl%
\url{https://doi.org/10.1145/3411813}
\showDOI{\tempurl}


\bibitem[Hao et~al\mbox{.}(2021)]%
        {10.1145/3447548.3467181}
\bibfield{author}{\bibinfo{person}{Qianyue Hao}, \bibinfo{person}{Fengli Xu},
  \bibinfo{person}{Lin Chen}, \bibinfo{person}{Pan Hui}, {and}
  \bibinfo{person}{Yong Li}.} \bibinfo{year}{2021}\natexlab{}.
\newblock \showarticletitle{Hierarchical Reinforcement Learning for Scarce
  Medical Resource Allocation with Imperfect Information}. In
  \bibinfo{booktitle}{\emph{Proceedings of the 27th ACM SIGKDD Conference on
  Knowledge Discovery \&amp; Data Mining}} (Virtual Event, Singapore)
  \emph{(\bibinfo{series}{KDD '21})}. \bibinfo{publisher}{Association for
  Computing Machinery}, \bibinfo{address}{New York, NY, USA},
  \bibinfo{pages}{2955–2963}.
\newblock
\showISBNx{9781450383325}
\urldef\tempurl%
\url{https://doi.org/10.1145/3447548.3467181}
\showDOI{\tempurl}


\bibitem[Iodice et~al\mbox{.}(2019)]%
        {iodice2019interoceptive}
\bibfield{author}{\bibinfo{person}{Pierpaolo Iodice},
  \bibinfo{person}{Giuseppina Porciello}, \bibinfo{person}{Ilaria Bufalari},
  \bibinfo{person}{Laura Barca}, {and} \bibinfo{person}{Giovanni Pezzulo}.}
  \bibinfo{year}{2019}\natexlab{}.
\newblock \showarticletitle{An interoceptive illusion of effort induced by
  false heart-rate feedback}.
\newblock \bibinfo{journal}{\emph{Proceedings of the National Academy of
  Sciences}} \bibinfo{volume}{116}, \bibinfo{number}{28}
  (\bibinfo{year}{2019}), \bibinfo{pages}{13897--13902}.
\newblock


\bibitem[Judd and Klingberg(2021)]%
        {judd2021training}
\bibfield{author}{\bibinfo{person}{Nicholas Judd} {and} \bibinfo{person}{Torkel
  Klingberg}.} \bibinfo{year}{2021}\natexlab{}.
\newblock \showarticletitle{Training spatial cognition enhances mathematical
  learning in a randomized study of 17,000 children}.
\newblock \bibinfo{journal}{\emph{Nature Human Behaviour}} \bibinfo{volume}{5},
  \bibinfo{number}{11} (\bibinfo{year}{2021}), \bibinfo{pages}{1548--1554}.
\newblock


\bibitem[Kosmyna and Maes(2019)]%
        {kosmyna2019attentivu}
\bibfield{author}{\bibinfo{person}{Nataliya Kosmyna} {and}
  \bibinfo{person}{Pattie Maes}.} \bibinfo{year}{2019}\natexlab{}.
\newblock \showarticletitle{Attentivu: An EEG-based closed-loop biofeedback
  system for real-time monitoring and improvement of engagement for
  personalized learning}.
\newblock \bibinfo{journal}{\emph{Sensors}} \bibinfo{volume}{19},
  \bibinfo{number}{23} (\bibinfo{year}{2019}), \bibinfo{pages}{5200}.
\newblock


\bibitem[Kosmyna et~al\mbox{.}(2018)]%
        {kosmyna2018attentivu}
\bibfield{author}{\bibinfo{person}{Nataliya Kosmyna}, \bibinfo{person}{Utkarsh
  Sarawgi}, {and} \bibinfo{person}{Pattie Maes}.}
  \bibinfo{year}{2018}\natexlab{}.
\newblock \showarticletitle{AttentivU: Evaluating the feasibility of
  biofeedback glasses to monitor and improve attention}. In
  \bibinfo{booktitle}{\emph{Proceedings of the 2018 ACM International Joint
  Conference and 2018 International Symposium on Pervasive and Ubiquitous
  Computing and Wearable Computers}}. \bibinfo{pages}{999--1005}.
\newblock


\bibitem[Liao et~al\mbox{.}(2020)]%
        {liao2020personalized}
\bibfield{author}{\bibinfo{person}{Peng Liao}, \bibinfo{person}{Kristjan
  Greenewald}, \bibinfo{person}{Predrag Klasnja}, {and} \bibinfo{person}{Susan
  Murphy}.} \bibinfo{year}{2020}\natexlab{}.
\newblock \showarticletitle{Personalized heartsteps: A reinforcement learning
  algorithm for optimizing physical activity}.
\newblock \bibinfo{journal}{\emph{Proceedings of the ACM on Interactive,
  Mobile, Wearable and Ubiquitous Technologies}} \bibinfo{volume}{4},
  \bibinfo{number}{1} (\bibinfo{year}{2020}), \bibinfo{pages}{1--22}.
\newblock


\bibitem[Lin et~al\mbox{.}(2011)]%
        {lin2011spatial}
\bibfield{author}{\bibinfo{person}{Chin-Teng Lin}, \bibinfo{person}{Shi-An
  Chen}, \bibinfo{person}{Tien-Ting Chiu}, \bibinfo{person}{Hong-Zhang Lin},
  {and} \bibinfo{person}{Li-Wei Ko}.} \bibinfo{year}{2011}\natexlab{}.
\newblock \showarticletitle{Spatial and temporal EEG dynamics of dual-task
  driving performance}.
\newblock \bibinfo{journal}{\emph{Journal of neuroengineering and
  rehabilitation}} \bibinfo{volume}{8}, \bibinfo{number}{1}
  (\bibinfo{year}{2011}), \bibinfo{pages}{1--13}.
\newblock


\bibitem[MacLean et~al\mbox{.}(2013)]%
        {maclean2013moodwings}
\bibfield{author}{\bibinfo{person}{Diana MacLean}, \bibinfo{person}{Asta
  Roseway}, {and} \bibinfo{person}{Mary Czerwinski}.}
  \bibinfo{year}{2013}\natexlab{}.
\newblock \showarticletitle{MoodWings: a wearable biofeedback device for
  real-time stress intervention}. In \bibinfo{booktitle}{\emph{Proceedings of
  the 6th international conference on PErvasive Technologies Related to
  Assistive Environments}}. \bibinfo{pages}{1--8}.
\newblock


\bibitem[Mayer(2003)]%
        {mayer2003causes}
\bibfield{author}{\bibinfo{person}{Richard~E Mayer}.}
  \bibinfo{year}{2003}\natexlab{}.
\newblock \showarticletitle{What causes individual differences in cognitive
  performance}.
\newblock \bibinfo{journal}{\emph{The psychology of abilities, competencies,
  and expertise}} (\bibinfo{year}{2003}), \bibinfo{pages}{263--273}.
\newblock


\bibitem[Meschtscherjakov et~al\mbox{.}(2011)]%
        {10.1145/2037373.2037501}
\bibfield{author}{\bibinfo{person}{Alexander Meschtscherjakov},
  \bibinfo{person}{Christiane Moser}, \bibinfo{person}{Manfred Tscheligi},
  {and} \bibinfo{person}{Erika Reponen}.} \bibinfo{year}{2011}\natexlab{}.
\newblock \showarticletitle{Mobile Work Efficiency: Enhancing Workflows with
  Mobile Devices}. In \bibinfo{booktitle}{\emph{Proceedings of the 13th
  International Conference on Human Computer Interaction with Mobile Devices
  and Services}} (Stockholm, Sweden) \emph{(\bibinfo{series}{MobileHCI '11})}.
  \bibinfo{publisher}{Association for Computing Machinery},
  \bibinfo{address}{New York, NY, USA}, \bibinfo{pages}{737–740}.
\newblock
\showISBNx{9781450305419}
\urldef\tempurl%
\url{https://doi.org/10.1145/2037373.2037501}
\showDOI{\tempurl}


\bibitem[Mirjafari et~al\mbox{.}(2021)]%
        {mirjafari2021predicting}
\bibfield{author}{\bibinfo{person}{Shayan Mirjafari}, \bibinfo{person}{Hessam
  Bagherinezhad}, \bibinfo{person}{Subigya Nepal}, \bibinfo{person}{Gonzalo~J
  Martinez}, \bibinfo{person}{Koustuv Saha}, \bibinfo{person}{Mikio Obuchi},
  \bibinfo{person}{Pino~G Audia}, \bibinfo{person}{Nitesh~V Chawla},
  \bibinfo{person}{Anind~K Dey}, \bibinfo{person}{Aaron Striegel},
  {et~al\mbox{.}}} \bibinfo{year}{2021}\natexlab{}.
\newblock \showarticletitle{Predicting Job Performance Using Mobile Sensing}.
\newblock \bibinfo{journal}{\emph{IEEE Pervasive Computing}}
  \bibinfo{volume}{20}, \bibinfo{number}{4} (\bibinfo{year}{2021}),
  \bibinfo{pages}{43--51}.
\newblock


\bibitem[Moon et~al\mbox{.}(2022)]%
        {moon2022speeding}
\bibfield{author}{\bibinfo{person}{Hee-Seung Moon}, \bibinfo{person}{Seungwon
  Do}, \bibinfo{person}{Wonjae Kim}, \bibinfo{person}{Jiwon Seo},
  \bibinfo{person}{Minsuk Chang}, {and} \bibinfo{person}{Byungjoo Lee}.}
  \bibinfo{year}{2022}\natexlab{}.
\newblock \showarticletitle{Speeding up Inference with User Simulators through
  Policy Modulation}. In \bibinfo{booktitle}{\emph{CHI Conference on Human
  Factors in Computing Systems}}. \bibinfo{pages}{1--21}.
\newblock


\bibitem[Moore and Tenney(2012)]%
        {moore2012time}
\bibfield{author}{\bibinfo{person}{Don~A Moore} {and}
  \bibinfo{person}{Elizabeth~R Tenney}.} \bibinfo{year}{2012}\natexlab{}.
\newblock \showarticletitle{Time pressure, performance, and productivity}.
\newblock In \bibinfo{booktitle}{\emph{Looking back, moving forward: A review
  of group and team-based research}}. Vol.~\bibinfo{volume}{15}.
  \bibinfo{publisher}{Emerald Group Publishing Limited},
  \bibinfo{pages}{305--326}.
\newblock


\bibitem[M{\"u}ller and Schumann(2011)]%
        {muller2011drugs}
\bibfield{author}{\bibinfo{person}{Christian~P M{\"u}ller} {and}
  \bibinfo{person}{Gunter Schumann}.} \bibinfo{year}{2011}\natexlab{}.
\newblock \showarticletitle{Drugs as instruments: a new framework for
  non-addictive psychoactive drug use}.
\newblock \bibinfo{journal}{\emph{Behavioral and Brain Sciences}}
  \bibinfo{volume}{34}, \bibinfo{number}{6} (\bibinfo{year}{2011}),
  \bibinfo{pages}{293}.
\newblock


\bibitem[Nicholson et~al\mbox{.}(2007)]%
        {10.5555/1231091.1231096}
\bibfield{author}{\bibinfo{person}{Darren Nicholson}, \bibinfo{person}{Diane
  Hamilton}, {and} \bibinfo{person}{Daniel McFarland}.}
  \bibinfo{year}{2007}\natexlab{}.
\newblock \showarticletitle{Leveraging Learning Styles to Improve Student
  Learning: The Interactive Learning Model and Learning Combination Inventory}.
\newblock \bibinfo{journal}{\emph{J. Comput. Sci. Coll.}} \bibinfo{volume}{22},
  \bibinfo{number}{6} (\bibinfo{date}{jun} \bibinfo{year}{2007}),
  \bibinfo{pages}{8–17}.
\newblock
\showISSN{1937-4771}


\bibitem[Nip et~al\mbox{.}(2018)]%
        {10.1145/3159450.3159500}
\bibfield{author}{\bibinfo{person}{Terence Nip}, \bibinfo{person}{Elsa~L.
  Gunter}, \bibinfo{person}{Geoffrey~L. Herman}, \bibinfo{person}{Jason~W.
  Morphew}, {and} \bibinfo{person}{Matthew West}.}
  \bibinfo{year}{2018}\natexlab{}.
\newblock \showarticletitle{Using a Computer-Based Testing Facility to Improve
  Student Learning in a Programming Languages and Compilers Course}. In
  \bibinfo{booktitle}{\emph{Proceedings of the 49th ACM Technical Symposium on
  Computer Science Education}} (Baltimore, Maryland, USA)
  \emph{(\bibinfo{series}{SIGCSE '18})}. \bibinfo{publisher}{Association for
  Computing Machinery}, \bibinfo{address}{New York, NY, USA},
  \bibinfo{pages}{568–573}.
\newblock
\showISBNx{9781450351034}
\urldef\tempurl%
\url{https://doi.org/10.1145/3159450.3159500}
\showDOI{\tempurl}


\bibitem[Noti et~al\mbox{.}(2016)]%
        {noti2016behavior}
\bibfield{author}{\bibinfo{person}{Gali Noti}, \bibinfo{person}{Effi Levi},
  \bibinfo{person}{Yoav Kolumbus}, {and} \bibinfo{person}{Amit Daniely}.}
  \bibinfo{year}{2016}\natexlab{}.
\newblock \showarticletitle{Behavior-based machine-learning: A hybrid approach
  for predicting human decision making}.
\newblock \bibinfo{journal}{\emph{arXiv preprint arXiv:1611.10228}}
  (\bibinfo{year}{2016}).
\newblock


\bibitem[Panigrahi(2017)]%
        {10.1145/3084381.3084434}
\bibfield{author}{\bibinfo{person}{Ritanjali Panigrahi}.}
  \bibinfo{year}{2017}\natexlab{}.
\newblock \showarticletitle{Online Learning: Improving the Learning Outcomes}.
  In \bibinfo{booktitle}{\emph{Proceedings of the 2017 ACM SIGMIS Conference on
  Computers and People Research}} (Bangalore, India)
  \emph{(\bibinfo{series}{SIGMIS-CPR '17})}. \bibinfo{publisher}{Association
  for Computing Machinery}, \bibinfo{address}{New York, NY, USA},
  \bibinfo{pages}{203–204}.
\newblock
\showISBNx{9781450350372}
\urldef\tempurl%
\url{https://doi.org/10.1145/3084381.3084434}
\showDOI{\tempurl}


\bibitem[Park and Lee(2020)]%
        {park2020intermittent}
\bibfield{author}{\bibinfo{person}{Eunji Park} {and} \bibinfo{person}{Byungjoo
  Lee}.} \bibinfo{year}{2020}\natexlab{}.
\newblock \showarticletitle{An intermittent click planning model}. In
  \bibinfo{booktitle}{\emph{Proceedings of the 2020 CHI Conference on Human
  Factors in Computing Systems}}. \bibinfo{pages}{1--13}.
\newblock


\bibitem[Paszke et~al\mbox{.}(2019)]%
        {NEURIPS2019_9015}
\bibfield{author}{\bibinfo{person}{Adam Paszke}, \bibinfo{person}{Sam Gross},
  \bibinfo{person}{Francisco Massa}, \bibinfo{person}{Adam Lerer},
  \bibinfo{person}{James Bradbury}, \bibinfo{person}{Gregory Chanan},
  \bibinfo{person}{Trevor Killeen}, \bibinfo{person}{Zeming Lin},
  \bibinfo{person}{Natalia Gimelshein}, \bibinfo{person}{Luca Antiga},
  \bibinfo{person}{Alban Desmaison}, \bibinfo{person}{Andreas Kopf},
  \bibinfo{person}{Edward Yang}, \bibinfo{person}{Zachary DeVito},
  \bibinfo{person}{Martin Raison}, \bibinfo{person}{Alykhan Tejani},
  \bibinfo{person}{Sasank Chilamkurthy}, \bibinfo{person}{Benoit Steiner},
  \bibinfo{person}{Lu Fang}, \bibinfo{person}{Junjie Bai}, {and}
  \bibinfo{person}{Soumith Chintala}.} \bibinfo{year}{2019}\natexlab{}.
\newblock \showarticletitle{PyTorch: An Imperative Style, High-Performance Deep
  Learning Library}.
\newblock In \bibinfo{booktitle}{\emph{Advances in Neural Information
  Processing Systems 32}}. \bibinfo{publisher}{Curran Associates, Inc.},
  \bibinfo{pages}{8024--8035}.
\newblock
\urldef\tempurl%
\url{http://papers.neurips.cc/paper/9015-pytorch-an-imperative-style-high-performance-deep-learning-library.pdf}
\showURL{%
\tempurl}


\bibitem[Pedregosa et~al\mbox{.}(2011)]%
        {scikit-learn}
\bibfield{author}{\bibinfo{person}{F. Pedregosa}, \bibinfo{person}{G.
  Varoquaux}, \bibinfo{person}{A. Gramfort}, \bibinfo{person}{V. Michel},
  \bibinfo{person}{B. Thirion}, \bibinfo{person}{O. Grisel},
  \bibinfo{person}{M. Blondel}, \bibinfo{person}{P. Prettenhofer},
  \bibinfo{person}{R. Weiss}, \bibinfo{person}{V. Dubourg}, \bibinfo{person}{J.
  Vanderplas}, \bibinfo{person}{A. Passos}, \bibinfo{person}{D. Cournapeau},
  \bibinfo{person}{M. Brucher}, \bibinfo{person}{M. Perrot}, {and}
  \bibinfo{person}{E. Duchesnay}.} \bibinfo{year}{2011}\natexlab{}.
\newblock \showarticletitle{Scikit-learn: Machine Learning in {P}ython}.
\newblock \bibinfo{journal}{\emph{Journal of Machine Learning Research}}
  \bibinfo{volume}{12} (\bibinfo{year}{2011}), \bibinfo{pages}{2825--2830}.
\newblock


\bibitem[Peterson et~al\mbox{.}(2018)]%
        {peterson2018evaluating}
\bibfield{author}{\bibinfo{person}{Joshua~C Peterson},
  \bibinfo{person}{Joshua~T Abbott}, {and} \bibinfo{person}{Thomas~L
  Griffiths}.} \bibinfo{year}{2018}\natexlab{}.
\newblock \showarticletitle{Evaluating (and improving) the correspondence
  between deep neural networks and human representations}.
\newblock \bibinfo{journal}{\emph{Cognitive science}} \bibinfo{volume}{42},
  \bibinfo{number}{8} (\bibinfo{year}{2018}), \bibinfo{pages}{2648--2669}.
\newblock


\bibitem[Peterson et~al\mbox{.}(2021)]%
        {peterson2021using}
\bibfield{author}{\bibinfo{person}{Joshua~C Peterson}, \bibinfo{person}{David~D
  Bourgin}, \bibinfo{person}{Mayank Agrawal}, \bibinfo{person}{Daniel
  Reichman}, {and} \bibinfo{person}{Thomas~L Griffiths}.}
  \bibinfo{year}{2021}\natexlab{}.
\newblock \showarticletitle{Using large-scale experiments and machine learning
  to discover theories of human decision-making}.
\newblock \bibinfo{journal}{\emph{Science}} \bibinfo{volume}{372},
  \bibinfo{number}{6547} (\bibinfo{year}{2021}), \bibinfo{pages}{1209--1214}.
\newblock


\bibitem[Plonsky et~al\mbox{.}(2017)]%
        {plonsky2017psychological}
\bibfield{author}{\bibinfo{person}{Ori Plonsky}, \bibinfo{person}{Ido Erev},
  \bibinfo{person}{Tamir Hazan}, {and} \bibinfo{person}{Moshe Tennenholtz}.}
  \bibinfo{year}{2017}\natexlab{}.
\newblock \showarticletitle{Psychological forest: Predicting human behavior}.
  In \bibinfo{booktitle}{\emph{Thirty-first AAAI conference on artificial
  intelligence}}.
\newblock


\bibitem[Pruettikomon and Louhapensang(2018)]%
        {pruettikomon2018study}
\bibfield{author}{\bibinfo{person}{Soraj Pruettikomon} {and}
  \bibinfo{person}{Chaturong Louhapensang}.} \bibinfo{year}{2018}\natexlab{}.
\newblock \showarticletitle{A study and development of workplace facilities and
  working environment to increase the work efficiency of persons with
  disabilities: A Case Study Of Major Retail And Wholesale Companies in
  Bangkok}.
\newblock \bibinfo{journal}{\emph{The Scientific World Journal}}
  \bibinfo{volume}{2018} (\bibinfo{year}{2018}).
\newblock


\bibitem[Raffin et~al\mbox{.}(2021)]%
        {stable-baselines3}
\bibfield{author}{\bibinfo{person}{Antonin Raffin}, \bibinfo{person}{Ashley
  Hill}, \bibinfo{person}{Adam Gleave}, \bibinfo{person}{Anssi Kanervisto},
  \bibinfo{person}{Maximilian Ernestus}, {and} \bibinfo{person}{Noah Dormann}.}
  \bibinfo{year}{2021}\natexlab{}.
\newblock \showarticletitle{Stable-Baselines3: Reliable Reinforcement Learning
  Implementations}.
\newblock \bibinfo{journal}{\emph{Journal of Machine Learning Research}}
  \bibinfo{volume}{22}, \bibinfo{number}{268} (\bibinfo{year}{2021}),
  \bibinfo{pages}{1--8}.
\newblock
\urldef\tempurl%
\url{http://jmlr.org/papers/v22/20-1364.html}
\showURL{%
\tempurl}


\bibitem[Ratcliff and McKoon(2008)]%
        {ratcliff2008diffusion}
\bibfield{author}{\bibinfo{person}{Roger Ratcliff} {and} \bibinfo{person}{Gail
  McKoon}.} \bibinfo{year}{2008}\natexlab{}.
\newblock \showarticletitle{The diffusion decision model: theory and data for
  two-choice decision tasks}.
\newblock \bibinfo{journal}{\emph{Neural computation}} \bibinfo{volume}{20},
  \bibinfo{number}{4} (\bibinfo{year}{2008}), \bibinfo{pages}{873--922}.
\newblock


\bibitem[Schulman et~al\mbox{.}(2017)]%
        {Schulman2017ProximalPO}
\bibfield{author}{\bibinfo{person}{John Schulman}, \bibinfo{person}{Filip
  Wolski}, \bibinfo{person}{Prafulla Dhariwal}, \bibinfo{person}{Alec Radford},
  {and} \bibinfo{person}{Oleg Klimov}.} \bibinfo{year}{2017}\natexlab{}.
\newblock \showarticletitle{Proximal Policy Optimization Algorithms}.
\newblock \bibinfo{journal}{\emph{ArXiv}}  \bibinfo{volume}{abs/1707.06347}
  (\bibinfo{year}{2017}).
\newblock


\bibitem[Scurto et~al\mbox{.}(2021)]%
        {scurto2021designing}
\bibfield{author}{\bibinfo{person}{Hugo Scurto}, \bibinfo{person}{Bavo~Van
  Kerrebroeck}, \bibinfo{person}{Baptiste Caramiaux}, {and}
  \bibinfo{person}{Fr{\'e}d{\'e}ric Bevilacqua}.}
  \bibinfo{year}{2021}\natexlab{}.
\newblock \showarticletitle{Designing deep reinforcement learning for human
  parameter exploration}.
\newblock \bibinfo{journal}{\emph{ACM Transactions on Computer-Human
  Interaction (TOCHI)}} \bibinfo{volume}{28}, \bibinfo{number}{1}
  (\bibinfo{year}{2021}), \bibinfo{pages}{1--35}.
\newblock


\bibitem[Shammas(2019)]%
        {shammas2019simple}
\bibfield{author}{\bibinfo{person}{Michael Shammas}.}
  \bibinfo{year}{2019}\natexlab{}.
\newblock \showarticletitle{Why a Simple Time-Management System Can
  Revolutionize How You Work—and Live}.
\newblock \bibinfo{journal}{\emph{Shammas, Michael." Why a Simple
  Time-Management System Can Revolutionize How You Work—And Live." Medium}}
  (\bibinfo{year}{2019}).
\newblock


\bibitem[Silva et~al\mbox{.}(2013)]%
        {silva2013benefits}
\bibfield{author}{\bibinfo{person}{Ana~R Silva}, \bibinfo{person}{Salom{\'e}
  Pinho}, \bibinfo{person}{Lu{\'\i}s~M Macedo}, {and} \bibinfo{person}{Chris~J
  Moulin}.} \bibinfo{year}{2013}\natexlab{}.
\newblock \showarticletitle{Benefits of SenseCam review on neuropsychological
  test performance}.
\newblock \bibinfo{journal}{\emph{American journal of preventive medicine}}
  \bibinfo{volume}{44}, \bibinfo{number}{3} (\bibinfo{year}{2013}),
  \bibinfo{pages}{302--307}.
\newblock


\bibitem[Singh et~al\mbox{.}(2020)]%
        {singh2020end}
\bibfield{author}{\bibinfo{person}{Pulkit Singh}, \bibinfo{person}{Joshua~C
  Peterson}, \bibinfo{person}{Ruairidh~M Battleday}, {and}
  \bibinfo{person}{Thomas~L Griffiths}.} \bibinfo{year}{2020}\natexlab{}.
\newblock \showarticletitle{End-to-end deep prototype and exemplar models for
  predicting human behavior}.
\newblock \bibinfo{journal}{\emph{arXiv preprint arXiv:2007.08723}}
  (\bibinfo{year}{2020}).
\newblock


\bibitem[Slobounov et~al\mbox{.}(2000)]%
        {slobounov2000neurophysiological}
\bibfield{author}{\bibinfo{person}{SM Slobounov}, \bibinfo{person}{K Fukada},
  \bibinfo{person}{R Simon}, \bibinfo{person}{M Rearick}, {and}
  \bibinfo{person}{W Ray}.} \bibinfo{year}{2000}\natexlab{}.
\newblock \showarticletitle{Neurophysiological and behavioral indices of time
  pressure effects on visuomotor task performance}.
\newblock \bibinfo{journal}{\emph{Cognitive Brain Research}}
  \bibinfo{volume}{9}, \bibinfo{number}{3} (\bibinfo{year}{2000}),
  \bibinfo{pages}{287--298}.
\newblock


\bibitem[Spielberger et~al\mbox{.}(1971)]%
        {spielberger1971state}
\bibfield{author}{\bibinfo{person}{Charles~D Spielberger},
  \bibinfo{person}{Fernando Gonzalez-Reigosa}, \bibinfo{person}{Angel
  Martinez-Urrutia}, \bibinfo{person}{Luiz~FS Natalicio}, {and}
  \bibinfo{person}{Diana~S Natalicio}.} \bibinfo{year}{1971}\natexlab{}.
\newblock \showarticletitle{The state-trait anxiety inventory}.
\newblock \bibinfo{journal}{\emph{Revista Interamericana de
  Psicologia/Interamerican Journal of Psychology}} \bibinfo{volume}{5},
  \bibinfo{number}{3 \& 4} (\bibinfo{year}{1971}).
\newblock


\bibitem[Spoerer et~al\mbox{.}(2020)]%
        {spoerer2020recurrent}
\bibfield{author}{\bibinfo{person}{Courtney~J Spoerer}, \bibinfo{person}{Tim~C
  Kietzmann}, \bibinfo{person}{Johannes Mehrer}, \bibinfo{person}{Ian Charest},
  {and} \bibinfo{person}{Nikolaus Kriegeskorte}.}
  \bibinfo{year}{2020}\natexlab{}.
\newblock \showarticletitle{Recurrent neural networks can explain flexible
  trading of speed and accuracy in biological vision}.
\newblock \bibinfo{journal}{\emph{PLoS computational biology}}
  \bibinfo{volume}{16}, \bibinfo{number}{10} (\bibinfo{year}{2020}),
  \bibinfo{pages}{e1008215}.
\newblock


\bibitem[Steinerman(2010)]%
        {steinerman2010minding}
\bibfield{author}{\bibinfo{person}{Joshua~R Steinerman}.}
  \bibinfo{year}{2010}\natexlab{}.
\newblock \showarticletitle{Minding the aging brain: technology-enabled
  cognitive training for healthy elders}.
\newblock \bibinfo{journal}{\emph{Current Neurology and neuroscience reports}}
  \bibinfo{volume}{10}, \bibinfo{number}{5} (\bibinfo{year}{2010}),
  \bibinfo{pages}{374--380}.
\newblock


\bibitem[Steyvers et~al\mbox{.}(2019)]%
        {steyvers2019large}
\bibfield{author}{\bibinfo{person}{Mark Steyvers}, \bibinfo{person}{Guy~E
  Hawkins}, \bibinfo{person}{Frini Karayanidis}, {and} \bibinfo{person}{Scott~D
  Brown}.} \bibinfo{year}{2019}\natexlab{}.
\newblock \showarticletitle{A large-scale analysis of task switching practice
  effects across the lifespan}.
\newblock \bibinfo{journal}{\emph{Proceedings of the National Academy of
  Sciences}} \bibinfo{volume}{116}, \bibinfo{number}{36}
  (\bibinfo{year}{2019}), \bibinfo{pages}{17735--17740}.
\newblock


\bibitem[Szafir and Mutlu(2012)]%
        {10.1145/2207676.2207679}
\bibfield{author}{\bibinfo{person}{Daniel Szafir} {and} \bibinfo{person}{Bilge
  Mutlu}.} \bibinfo{year}{2012}\natexlab{}.
\newblock \showarticletitle{Pay Attention! Designing Adaptive Agents That
  Monitor and Improve User Engagement}. In
  \bibinfo{booktitle}{\emph{Proceedings of the SIGCHI Conference on Human
  Factors in Computing Systems}} (Austin, Texas, USA)
  \emph{(\bibinfo{series}{CHI '12})}. \bibinfo{publisher}{Association for
  Computing Machinery}, \bibinfo{address}{New York, NY, USA},
  \bibinfo{pages}{11–20}.
\newblock
\showISBNx{9781450310154}
\urldef\tempurl%
\url{https://doi.org/10.1145/2207676.2207679}
\showDOI{\tempurl}


\bibitem[Tatsuta et~al\mbox{.}(2017)]%
        {10.1145/3055635.3056616}
\bibfield{author}{\bibinfo{person}{Riki Tatsuta}, \bibinfo{person}{Dinh
  Thi~Dong Phuong}, \bibinfo{person}{Yusuke Kajiwara}, {and}
  \bibinfo{person}{Hiromitsu Shimakawa}.} \bibinfo{year}{2017}\natexlab{}.
\newblock \showarticletitle{Guidance of Farming Works to Improve Efficiency
  Considering Physical Behavior}. In \bibinfo{booktitle}{\emph{Proceedings of
  the 9th International Conference on Machine Learning and Computing}}
  (Singapore, Singapore) \emph{(\bibinfo{series}{ICMLC 2017})}.
  \bibinfo{publisher}{Association for Computing Machinery},
  \bibinfo{address}{New York, NY, USA}, \bibinfo{pages}{28–32}.
\newblock
\showISBNx{9781450348171}
\urldef\tempurl%
\url{https://doi.org/10.1145/3055635.3056616}
\showDOI{\tempurl}


\bibitem[Vouros(2022)]%
        {10.1145/3527448}
\bibfield{author}{\bibinfo{person}{George~A. Vouros}.}
  \bibinfo{year}{2022}\natexlab{}.
\newblock \showarticletitle{Explainable Deep Reinforcement Learning: State of
  the Art and Challenges}.
\newblock \bibinfo{journal}{\emph{ACM Comput. Surv.}} \bibinfo{volume}{55},
  \bibinfo{number}{5}, Article \bibinfo{articleno}{92} (\bibinfo{date}{dec}
  \bibinfo{year}{2022}), \bibinfo{numpages}{39}~pages.
\newblock
\showISSN{0360-0300}
\urldef\tempurl%
\url{https://doi.org/10.1145/3527448}
\showDOI{\tempurl}


\bibitem[Warnell et~al\mbox{.}(2018)]%
        {warnell2018deep}
\bibfield{author}{\bibinfo{person}{Garrett Warnell}, \bibinfo{person}{Nicholas
  Waytowich}, \bibinfo{person}{Vernon Lawhern}, {and} \bibinfo{person}{Peter
  Stone}.} \bibinfo{year}{2018}\natexlab{}.
\newblock \showarticletitle{Deep tamer: Interactive agent shaping in
  high-dimensional state spaces}. In \bibinfo{booktitle}{\emph{Proceedings of
  the AAAI conference on artificial intelligence}}, Vol.~\bibinfo{volume}{32}.
\newblock


\bibitem[Whittaker et~al\mbox{.}(2016)]%
        {whittaker2016don}
\bibfield{author}{\bibinfo{person}{Steve Whittaker}, \bibinfo{person}{Vaiva
  Kalnikaite}, \bibinfo{person}{Victoria Hollis}, {and} \bibinfo{person}{Andrew
  Guydish}.} \bibinfo{year}{2016}\natexlab{}.
\newblock \showarticletitle{'Don't Waste My Time' Use of Time Information
  Improves Focus}. In \bibinfo{booktitle}{\emph{Proceedings of the 2016 CHI
  Conference on Human Factors in Computing Systems}}.
  \bibinfo{pages}{1729--1738}.
\newblock


\bibitem[Xu and Zhang(2023)]%
        {xu2023modeling}
\bibfield{author}{\bibinfo{person}{Songlin Xu} {and} \bibinfo{person}{Xinyu
  Zhang}.} \bibinfo{year}{2023}\natexlab{}.
\newblock \showarticletitle{Modeling Human Cognition with a Hybrid Deep
  Reinforcement Learning Agent}.
\newblock \bibinfo{journal}{\emph{arXiv preprint arXiv:2301.06216}}
  (\bibinfo{year}{2023}).
\newblock


\bibitem[Yang et~al\mbox{.}(2019)]%
        {yang2019task}
\bibfield{author}{\bibinfo{person}{Guangyu~Robert Yang},
  \bibinfo{person}{Madhura~R Joglekar}, \bibinfo{person}{H~Francis Song},
  \bibinfo{person}{William~T Newsome}, {and} \bibinfo{person}{Xiao-Jing Wang}.}
  \bibinfo{year}{2019}\natexlab{}.
\newblock \showarticletitle{Task representations in neural networks trained to
  perform many cognitive tasks}.
\newblock \bibinfo{journal}{\emph{Nature neuroscience}} \bibinfo{volume}{22},
  \bibinfo{number}{2} (\bibinfo{year}{2019}), \bibinfo{pages}{297--306}.
\newblock


\bibitem[Zhai et~al\mbox{.}(2004)]%
        {zhai2004speed}
\bibfield{author}{\bibinfo{person}{Shumin Zhai}, \bibinfo{person}{Jing Kong},
  {and} \bibinfo{person}{Xiangshi Ren}.} \bibinfo{year}{2004}\natexlab{}.
\newblock \showarticletitle{Speed--accuracy tradeoff in Fitts’ law tasks—on
  the equivalency of actual and nominal pointing precision}.
\newblock \bibinfo{journal}{\emph{International journal of human-computer
  studies}} \bibinfo{volume}{61}, \bibinfo{number}{6} (\bibinfo{year}{2004}),
  \bibinfo{pages}{823--856}.
\newblock


\bibitem[Zur and Breznitz(1981)]%
        {zur1981effect}
\bibfield{author}{\bibinfo{person}{Hasida~Ben Zur} {and}
  \bibinfo{person}{Shlomo~J Breznitz}.} \bibinfo{year}{1981}\natexlab{}.
\newblock \showarticletitle{The effect of time pressure on risky choice
  behavior}.
\newblock \bibinfo{journal}{\emph{Acta Psychologica}} \bibinfo{volume}{47},
  \bibinfo{number}{2} (\bibinfo{year}{1981}), \bibinfo{pages}{89--104}.
\newblock


\end{thebibliography}

\end{document}